\documentclass{optica-article}

\journal{opticajournal}

\articletype{Research Article}

\usepackage{lineno}

\usepackage{graphicx}
\usepackage{dcolumn}
\usepackage{bm}

\usepackage{booktabs} 
\usepackage{multirow}
\usepackage{tabularx}
\usepackage{comment}
\usepackage{xcolor}
\usepackage{alltt}
\usepackage{listings}
\usepackage{rotating,tablefootnote}

\usepackage{bibunits}

\defaultbibliographystyle{opticajnl} 
\defaultbibliography{refs_optica} 
\makeatletter
\let\optica@orig@bibliographystyle\bibliographystyle
\renewcommand{\bibliographystyle}[1]{} 
\makeatother
\usepackage{hyperref}

\begin{document}
\begin{bibunit}

\title{AI Agents for Photonic Integrated Circuit Design Automation}

\author{%
  Ankita Sharma\authormark{1,2,$\dagger$,*},
  Yuqi Fu\authormark{1,$\dagger$,*},
  Vahid Ansari\authormark{3,$\dagger$},
  Rishabh Iyer\authormark{1},
  Fiona Kuang\authormark{1},
  Kashish Mistry\authormark{1},
  Raisa Islam Aishy\authormark{1},
  Sara Ahmad\authormark{1},
  Joaquin Matres\authormark{3},
  Dirk R. Englund\authormark{4,5},
  and Joyce K. S. Poon\authormark{1,*}
}

\address{%
\authormark{1}University of Toronto, 10 King’s College Road, Toronto, Ontario M5S 3G4, Canada\\
\authormark{2}Max Planck Institute of Microstructure Physics, Weinberg 2, 06120 Halle, Germany\\
\authormark{3}GDSFactory\\
\authormark{4}Research Laboratory of Electronics, Massachusetts Institute of Technology, 50 Vassar Street, Cambridge, MA 02139, USA\\
\authormark{5}Axiomatic\_AI Inc., 501 Massachusetts Ave., Cambridge, MA 02139, USA\\
\authormark{$\dagger$}These authors contributed equally to this work.
}

\email{\authormark{*}Corresponding authors: ank.sharma@mail.utoronto.ca; tony.fu@mail.utoronto.ca; joyce.poon@utoronto.ca}

\begin{abstract*}
We present Photonics Intelligent Design and Optimization (PhIDO), a multi-agent framework that converts natural-language photonic integrated circuit (PIC) design requests into layout mask files. We compare 7 reasoning large language models for PhIDO using a testbench of 102 design descriptions that ranged from single devices to 112-component PICs. The success rate for single-device designs was up to 91\%. For design queries with $\leq 15$ components, \texttt{o1}, \texttt{Gemini-2.5-pro}, and \texttt{Claude Opus 4} achieved the highest end-to-end pass@5 success rates of $\sim 57\%$, with \texttt{Gemini-2.5-pro} requiring the fewest output tokens and lowest cost.  The next steps toward autonomous PIC development include standardized knowledge representations, expanded datasets, extended verification, and robotic automation.

\end{abstract*}

\section{Introduction}

Generative artificial intelligence (AI) is transforming how scientific tasks are approached. Large language models (LLMs), based on generative pretrained transformer (GPT) architectures and trained on broad, heterogeneous datasets \cite{Minaee2024, Vaswani2017, Brown2020}, are capable of parsing unstructured inputs, conducting multi-stage reasoning, and interfacing with domain-specific toolchains.  LLM-based AI agents refined through instruction tuning and tool integration can autonomously complete complex tasks without continuous human oversight \cite{ouyang2022training, wei2022chain, yao2022react, schick2023toolformer, chung2024scaling, bai2022constitutional, wang2023self}. Such agents are being explored for scientific domains, such as chemistry, material science, and quantum computing \cite{wang2023scientific, boiko2023autonomous, lu2024aiscientistfullyautomated, gottweis2025aicoscientist, cao2025agentsselfdrivinglaboratoriesapplied}, where they assist scientists in generating hypothesis, planning experiments, surveying the literature, and evaluating research novelty. When combined with robotics, agents enable ``self-driving labs'' that autonomously complete experiments \cite{abolhasani2023rise,szymanski2023autonomous,tom2024self,Ghafarollahi2025, uddin2025aidrivenroboticsfreespaceoptics}. In electronic design automation (EDA), design assistants based on LLMs help automate hardware description language  synthesis and verification \cite{chang2023chipgptfarnaturallanguage, zhong2023llm4edaemergingprogresslarge,thakur2023verigenlargelanguagemodel,liu2024chipnemodomainadaptedllmschip, he2024chateda, ghimire2025hardwaredesignsecurityneeds,Xie2025FoundationAI}. They can also be integrated with optimizers and reinforcement learning for complete design flows \cite{mirhoseini2021graph,goldie2024addendum,Zheng2019NOC,Settaluri2022}. 

Despite these developments, photonic integrated circuit (PIC) design lacks robust automation tools and the use of AI agents  remains underexplored. Today, photonic design relies heavily on manual workflows and requires co-design across multiple physical domains (e.g., electromagnetic, semiconductor, thermal, electronic). The design process typically involves specification, selection and configuration of components, schematic construction, layout generation, and simulation-based validation \cite{Bogaerts2018}. Past efforts to apply LLMs to PIC design have been limited to scripting assistance with human-in-the-loop generation of specific optical components \cite{li2023llmhelpsdesignoptimize} or to automatic generation of netlists  \cite{wu2025picbenchbenchmarkingllmsphotonic}. These efforts have not demonstrated an end-to-end design workflow that starts from a natural language query and ends with a layout. 

Here, we present Photonics Intelligent Design and Optimization (PhIDO), an LLM-based agentic framework for PIC design automation, and datasets for AI-driven PIC design benchmarking. PhIDO interprets natural language input to extract design intent, generates circuit templates, constructs parametric netlists, simulates devices, and produces mask files in GDSII format. In this proof of concept work, our primary endpoint is structural validity rather than performance using a commercial process design kit. We used PhIDO to compare the performance of 7 LLMs for PIC design using a set of 102 natural language design requests ranging from single devices to large PICs comprising up to 112 devices. \texttt{o1}, \texttt{Gemini-2.5-pro}, and \texttt{Claude Opus 4} were found to have the highest success rates for the workflow with differences in costs and inference time. To our knowledge, this is the first comparative study of LLMs for end-to-end PIC design synthesis, from natural language input to GDSII mask file. The architecture of PhIDO may be extensible to other scientific design problems beyond photonics, such as free-space optics, microfluidics, micro-electromechanical systems, and radio-frequency circuits, where natural language inputs need interpretation based on domain expertise, design constraints are physics-based, and outputs are structured layouts or simulations. 

The methodology and results presented here are based on available LLMs between 2024 and mid-2025. Because generative AI is advancing at an extraordinary pace, the results should be viewed as a snapshot of present capabilities rather than a hard limit. We have made PhIDO open source and available for further development (see Data and Code Availability statement).

\section{PhIDO Architecture}

\subsection{Overview}

\begin{figure} 
    \centering
    \includegraphics[width=1\linewidth]{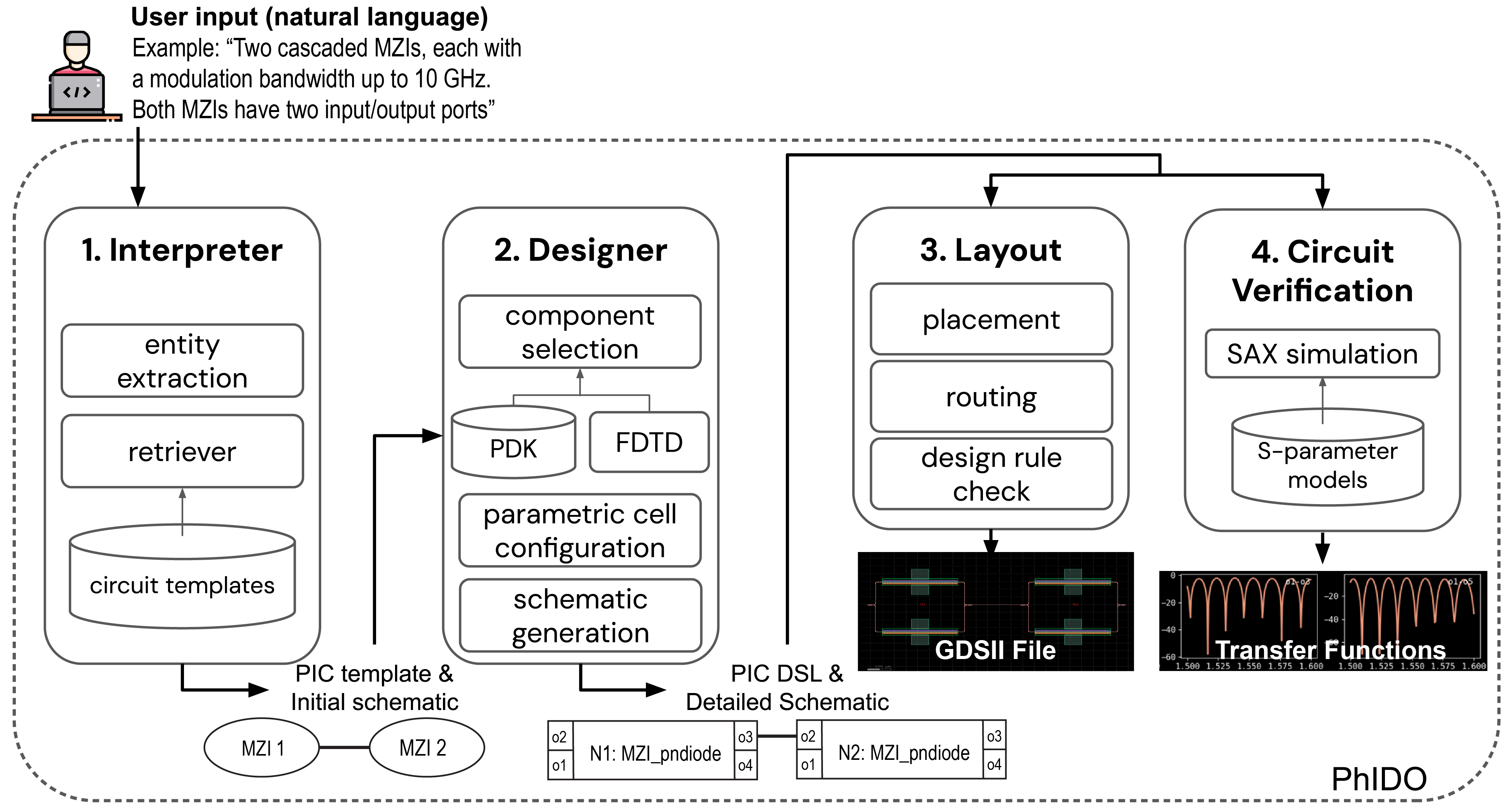}
    \caption{Overview of PhIDO. PhIDO is a multi-agent framework for PIC design automation. It generates GDSII layouts from natural language inputs in 4 stages: (1) Interpreter, an LLM-based agent that extracts entities and searches template databases to compose a draft circuit; (2) Designer, an LLM-based agent that generates a specific PIC design using components from a PDK; (3) Layout, an algorithmic module that produces the GDSII layout and performs placement and routing; and (4) Circuit verification, an algorithmic module that simulates the circuit using SAX. At each stage, users may inspect and modify outputs before proceeding to ensure design integrity.} 
    \label{fig:overview}
\end{figure}

PhIDO is structured as a modular platform with four main components as shown in Fig. \ref{fig:overview}: (1) Interpreter, (2) Designer, (3) Layout, and (4)  Circuit verification. Together, these modules translate free-form natural language inputs (i.e., prompts) into PIC layouts in GDSII file format. The Interpreter and Designer are agents, using instruction-tuned LLMs guided by in-context examples and chain-of-thought decomposition to break down the tasks \cite{wei2022chain}. Fine-tuning has been avoided because high-quality labeled PIC corpora are not available. Instead, each agent uses retrieval-augmented generation (RAG) \cite{LewisNEURIPS2020}, drawing on curated domain-specific knowledge, such as circuit templates and the information found in process design kits (PDKs) during inference. To improve reliability and mitigate hallucinations, outputs are verified through rule-based programmatic checks or self-refinement \cite{madaan2023selfrefineiterativerefinementselffeedback}. The Layout and Circuit verification stages are algorithmic modules integrated into the pipeline. The outputs of all four stages can be exposed to the user through a separate step-by-step workflow. The user may intervene and modify the pipeline's output at any step to ensure the integrity of input to the next step of the pipeline.

\begin{figure}
    \centering
    \includegraphics[width=1\linewidth]{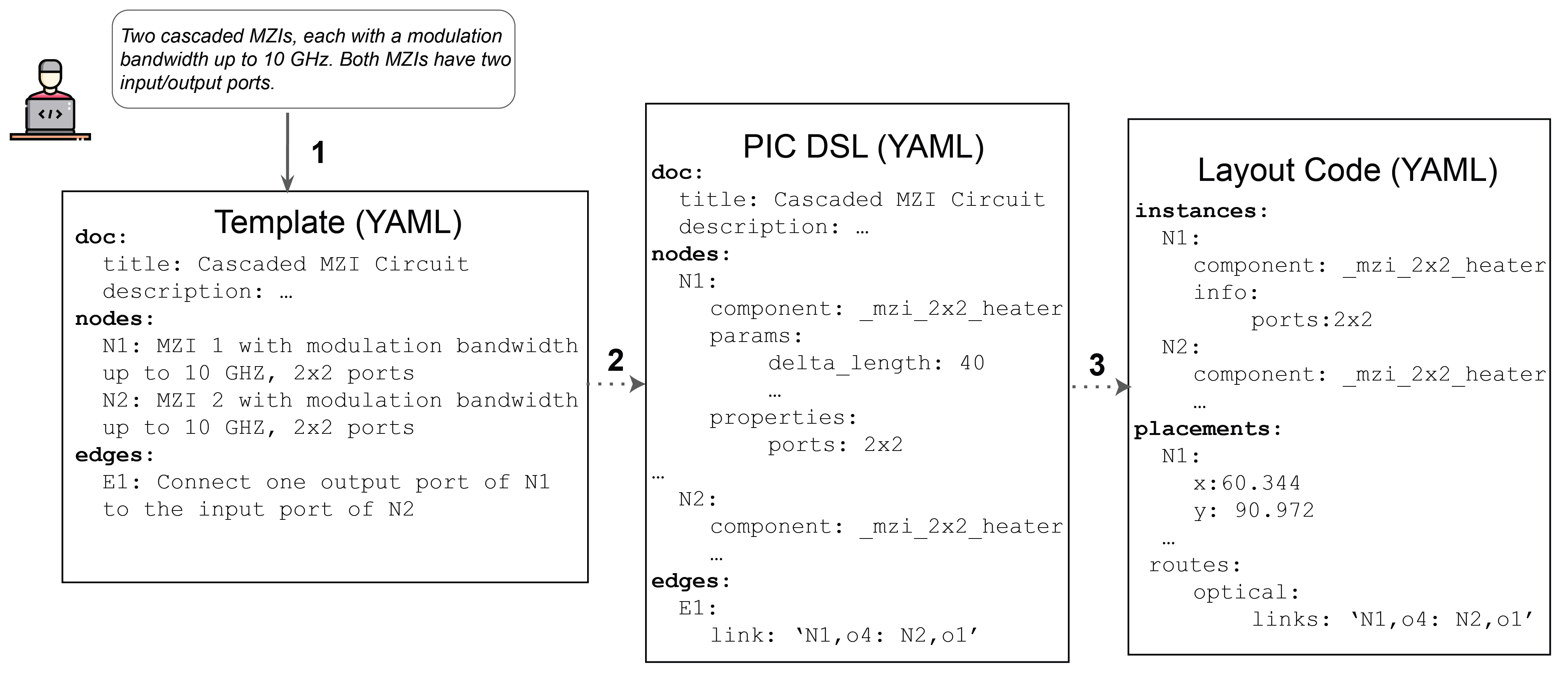}
 \caption{ DSL in PhIDO. The schematic shows an example of how natural language prompt is translated in layout code using the DSL. 1. At the output of the Interpreter, a PIC template is created as an intermediary representation of the user input. 2. At the output of the Designer, the PIC DSL, which is more specific to the design, is created from the PIC template. 3. The GDSFactory-compatible layout code is generated by the Layout module. These representations are in YAML.}
    \label{fig:dsl}
\end{figure}

A key aspect of PhIDO is the domain-specific language (DSL) that serves as an intermediate representation between natural language and technically specific description. Since much of the reasoning in PIC design is not formalized nor explicitly stated in user prompts, the DSL provides a machine-readable format for capturing design intent. The DSL is in YAML to be  compatible with GDSFactory, a widely used open-source PIC layout tool. It encodes component parameters, connectivity, and metadata such as PDK selection and target wavelengths. Figure \ref{fig:dsl} shows an example of how a natural language input is translated into DSL, which bridges informal specifications and formal PIC representations. The DSL schema defines components as nodes and interconnections as edges, along with their attributes. This structured representation enables more robust script generation, filters out extraneous information, and embeds physical constraints and design semantics critical to layout synthesis. 

\subsection{Agent configuration and system prompt design} \label{sec:config}

\begin{figure}
    \centering
    \includegraphics[width=1\linewidth]{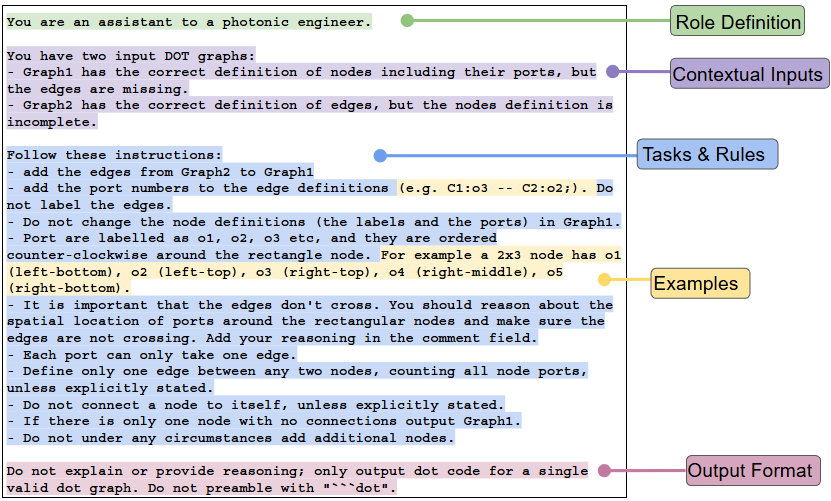}
    \caption{Annotated example of a system prompt used in  PhIDO. This example, from the Designer agent for schematic generation, highlights the five key attributes used to design prompts for task-specific reasoning.}
    \label{fig:prompt}
\end{figure}

The multi-agent workflow is guided by system prompts to ensure consistent and reproducible outputs. Each system prompt follows a structured format with five key elements: (1) role definition, specifying the agent’s function (e.g., a photonic chip layout developer); (2) contextual inputs, provided as structured data to ground the decision-making process (e.g., a JSON list of PDK components annotated with descriptive docstrings); (3) task instructions, outlining objectives and rule-based constraints that guide reasoning; (4) embedded examples, illustrating reasoning steps and expected outputs to promote consistency; and (5) output formatting requirements, enforcing standardized representations (e.g., YAML or DOT syntax) to enable seamless downstream processing. An example of a system prompt  with its key properties labeled is shown in Fig. \ref{fig:prompt}.


\subsubsection{Schema enforcement} \label{sec:Pydantic}

For stages that require  strict output structure for consistent parsing and downstream processing (i.e., entity extraction in the Interpreter agent and component selection in the Designer agent as described in Sections \ref{sec:Interpreter} and \ref{sec:Designer}), the outputs  were validated using Pydantic-enforced schema calls. A Pydantic model is a Python data class that defines a strict schema for the expected output (field names and types) and automatically validates that the response conforms to this schema \cite{pydantic}. OpenAI models natively support Pydantic validation via the OpenAI Python software design kit (SDK), but other models do not support SDK-level schema enforcement. While these models can be prompted to return structured JSON and validated post hoc, to ensure consistency in the workflow and schema enforcement, the structured tasks for non-Open AI models were routed through \texttt{GPT-4o} for Pydantic validation. A drawback to this approach is a mixing of models which can sometimes cause errors (see Section \ref{sec:sourcesoferrors} and Supplemental Document).

\subsection{Interpreter agent} \label{sec:Interpreter}

The Interpreter agent translates natural language into an intermediate PIC template and a high-level schematic that outlines the circuit topology. It mimics how a designer breaks down a design problem and reviews previous designs for inspiration. The PIC templates are abstract representations of the PIC DSL, capturing the structural layout of the circuit without detailed components or parameter values. Template generation has two main steps: (1) Entity extraction and (2) the Retriever.

In entity extraction, the agent identifies key design elements in the prompt such as functional blocks, connectivity, and performance constraints, then generates an initial schematic. First, a schema-constrained LLM uses a Pydantic-enforced prompt to extract a structured YAML component list from the natural language input (see Section \ref{sec:Pydantic}), identifying functional blocks, connectivity, and performance constraints. This YAML list is then passed to a second, schema-free LLM that generates an initial schematic representing the components and their relationships.

Afterwards, the Retriever searches a library of prior designs sourced from journal articles to find templates that match the parsed circuit. Users can either adopt a retrieved template or build one from scratch using the extracted entities. The YAML DSL represents components as nodes and optical links as edges, with additional attributes for PDK, wavelength band, and port types. An example of a generated PIC template is shown in Figure \ref{fig:dsl}(b).

\subsection{Designer agent} \label{sec:Designer}

The generated PIC template is passed to the Designer agent, which converts it into a fully specified circuit by composing the PIC DSL and generating a detailed schematic with port information. The agent carries out three main steps: (1) Component selection, (2) Parametric cell configuration, and (3) Schematic generation. 

First, the Designer agent maps the blocks in the template to candidate components in the model PDK that we have curated to test PhIDO and returns a ranked shortlist (see Fig. \ref{fig:PDKdsl}(a)). Future work can use a foundry-specific PDK within the bounds of confidentiality. In this component selection stage, each node in the PIC template is mapped to a candidate from the PDK. This matching process is implemented via a Pydantic-enforced LLM call (see Section \ref{sec:Pydantic}). Based on the intent of the design, the agent prioritizes functionality, followed by port structure, to produce a ranked list of matching candidates. Each match is qualitatively scored as ``exact'' (both functionality and port configuration match), ``partial'' (minor differences in functionality or port arrangement), or ``poor'' (significant mismatch). The user is presented with a ranked list of the top matches and can select from among the most relevant options.  The user has the option to choose a component among the list of top matches (Fig. \ref{fig:PDKdsl}(b)), and full automation is possible for unambiguous cases. 

Next, in the parameter configuration stage, the selected PCells are configured using geometric parameters parsed from the user input (Fig. \ref{fig:PDKdsl}(c)). These parameters may include specified waveguide widths, ring radii, grating periods, or other component-specific values. If the parameters deviate from their default values, the associated S-parameter models are automatically regenerated from the component layout using FDTD simulations. As a proof of concept, in a branch of the PhIDO repository, we integrate the Flexcompute Tidy3D finite difference time domain (FDTD) electromagnetic solver with automated optimization toward a figure of merit. This feature is not included in the main release due to compatibility issues between the GDSFactory version and Tidy3D FDTD solver but can be made available in future updates.

Lastly, in the schematic generation stage, the  agent completes the circuit by assigning explicit port-to-port connections using the initial schematic from the Interpret agent and the fully defined components from the PDK. While the initial schematic from the Interpreter specifies which components should connect, it lacks port-level detail. The Designer agent assigns edges by reasoning over component geometry, port orientation, and connectivity constraints, ensuring one connection per port, no self-loops, and minimal edge crossings. Each connection is annotated with specific port labels (e.g., C1:o2 -- C3:o5), resulting in a layout-aware schematic graph suitable for downstream synthesis and simulation. Because initially generated edge-to-port assignments can be error-prone, additional validation is introduced at this step. Generated schematics pass through a verification loop involving a second LLM to check syntax and planarity, followed by a crossing detection algorithm. If any edge crossings are found, the system updates the prompt context with the identified error and re-queries the agent to generate a corrected schematic. The finalized schematic is represented as a graph suitable for layout synthesis and downstream simulations.

\subsubsection{Generic PDK}\label{sec:PDK}

\begin{figure}
    \centering
    \includegraphics[width=1\linewidth]{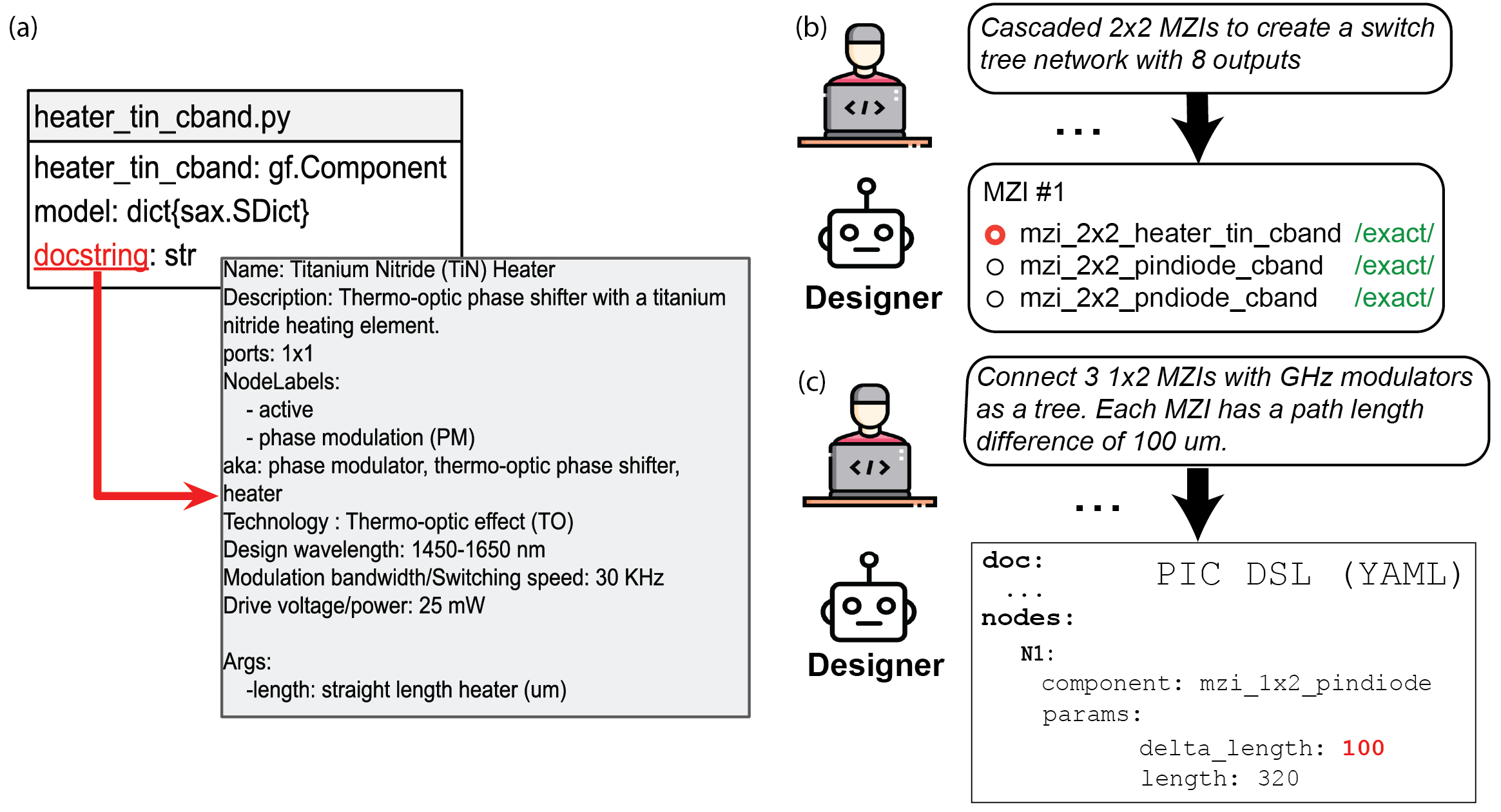}
    \caption{Designer agent interaction with the generic silicon photonics PDK in the PhIDO framework. (a) Example entry from the C-band PDK, illustrating structured annotations for a titanium nitride (TiN) heater. (b) The Designer agent searches over PDK docstrings to perform component selection, returning a ranked list of candidates based on functional and structural compatibility with the prompt intent. (c) The PIC DSL is constructed from a PIC template and populated with selected PDK components, which are then configured using geometric parameters extracted from the user prompt.}
    \label{fig:PDKdsl}
\end{figure}

We created a generic silicon photonics PDK centered at the telecom C-band (around 1550 nm). This PDK is not tied to any specific foundry and is free from confidentiality restrictions. It comprises 34 parametric cells (PCells) representing standard photonic building blocks, including waveguides, bends, directional couplers, and modulators. Each PCell is implemented as a GDSFactory object with an associated default S-parameter model for circuit-level simulations. PCells are annotated with standardized photonic-specific descriptors, covering functionality, optical port configuration, typical use cases, technology, and key performance parameters, to support retrieval in LLM-guided searches. Structured docstrings consolidate this information in a consistent format. An example of the level of labeling and structured docstring used for a titanium nitride (TiN) heater PCell is shown in Fig. \ref{fig:PDKdsl}(a).

\subsection{Layout \& Circuit verification agent}

The Layout agent generates the GDSII mask layout based on the circuit schematic produced by the Design agent. Component placement is performed using the DOT algorithm from the Graphviz library \cite{graphviz_dot}, with each graph node scaled to match the physical dimensions of the corresponding photonic component. These placement coordinates are translated into PhIDO’s photonic DSL and compiled into a GDSII file using GDSFactory. Optical routing is handled by GDSFactory’s built-in river router, which connects groups of parallel waveguides without crossings. The resulting layout undergoes design rule checks (DRC) to identify violations and is subsequently simulated using SAX circuit simulator for the user to evaluate the overall performance \cite{sax_photonics}.

To assess the layout pipeline independently, 118 test cases were constructed using an expanded PDK that included a broader range of component types and circuit sizes from the GDSFactory library. Each case was stored in YAML format, containing both the DOT schematic and the corresponding GDSFactory netlist. Using GDSFactory v8.16.0, 74 of the 118 circuits were successfully routed with default settings. Future releases of PhIDO will adopt updated versions of GDSFactory, as newer releases (e.g., v9.11.1) demonstrate improved routing success (see Supplemental Document).

\section{Testbench}

\subsection{Entries and complexity}

We evaluated the performance of PhIDO using a testbench of 102 natural language user prompts spanning a broad range of photonic design complexities. These prompts were grouped into four complexity levels. Table \ref{tab:levels} shows definition of the complexity levels along with the example prompts. Level 1 involves single components, while Level 4 are large PICs. These inputs varied in length and specificity and comprised manually written instructions, GPT-generated queries, and natural language formulations of circuits. Of the 102 prompts in the testbench, 23 (22.5\%) are Level 1, 10 (9.8\%) are Level 2, 60 (58.8\%) are Level 3, and 9 (8.8\%) are Level 4 (see Fig. \ref{fig:bench}). This testbench also includes 22 entries modified from  \cite{wu2025picbenchbenchmarkingllmsphotonic}. The testbench is available in the PhIDO repository on Github (see Data and Code Availability statement).


\begin{table}
\centering
\caption{Definition of the complexity levels in the testbench.} \label{tab:levels}
\begin{tabular}{@{}p{0.8cm}p{3cm}p{8.1cm}@{}}
\toprule
\textbf{Level} & \textbf{Component Count} & \textbf{Example Prompt} \\
\midrule
1 & 1  &
A $2\times 2$ MZI with a phase shifter in each arm for 1\,GHz modulation. \\
2 & 1 to 2  with a single connecting edge &
An $1\times 2$ MMI followed by a $2\times 2$ MMI. \\
3 & 3 to 15  &
Build a 4-channel wavelength demultiplexer by combining three two-channel demultiplexer blocks. Each two-channel block uses four cascaded Mach–Zehnder interferometers with path-length differences of 30\,\textmu m, 50\,\textmu m, 30\,\textmu m, and 50\,\textmu m, and incorporates $2\times 2$ MMIs as splitters and combiners. Arrange the first block to separate channels 1 and 2, the second block to separate channels 3 and 4, and then route all four outputs into the final block to produce four distinct outputs. The resulting device shall have four inputs and four outputs. \\
4 & 16 to 112 &
Design a $1\times 64$ Mach–Zehnder tree using six stages of $1\times 2$ MZIs. Each MZI has integrated thermo-optic phase shifters and MMI couplers. Use edge couplers at the single input and at each of the 64 outputs for fiber connection. Additionally, include a separate waveguide loopback structure, containing two edge couplers, connected to a $180^{\circ}$ bend. \\
\bottomrule
\end{tabular}
\end{table}
 

To improve the clarity in specifying circuit components and their connections within the testbench prompts, an initial benchmark evaluation was conducted using \texttt{gemini-1.5-pro}. This evaluation identified 22 problematic prompts that consistently failed during entity extraction or component specification stages. These ambiguous prompts were subsequently provided to \texttt{o1} for refinement. \texttt{o1} was explicitly tasked with identifying and clearly describing any ambiguities or underspecified aspects within the prompts, such as unclear component specifications or connectivity details, and recommending concrete improvements. \texttt{o1} then produced revised prompts incorporating these clarifications. For example, the original prompt  ``Cascaded $1\times2$ MZIs to create a $1\times16$ tree'' was refined to explicitly state the use of fifteen balanced $1\times 2$ Mach-Zehnder Interferometers arranged as a tree, with integrated phase shifters. Similarly, the prompt ``Design a $1\times16$ power splitter'' was improved by specifying the use of four stages of $1\times 2$ MMI couplers, ensuring each output port receives equal input power. The refined prompts were subsequently reviewed, validated, and, where necessary, further adjusted by the authors. Preliminary results indicate that using the \texttt{o1}-refined testbench prompts increased the absolute success rate by 3\% using \texttt{gemini-1.5-pro}. 

\subsection{Large language models}

We compared seven LLMs for PhIDO: OpenAI’s \texttt{o1} \cite{openai_o1} and \texttt{o3-mini} \cite{openai_o3mini}, Google’s \texttt{Gemini 1.5-pro} \cite{geminiteam2024gemini15unlockingmultimodal} and \texttt{Gemini 2.5-pro} \cite{geminiteam2025gemini25pushingfrontier}, Anthropic’s \texttt{Claude Opus 4} \cite{anthropic_claude_opus4, anthropic_claude_announce}, DeepSeek’s \texttt{R1} \cite{deepseekai2025deepseekr1incentivizingreasoningcapability}, and Nvidia’s \texttt{llama\allowbreak-3.1\allowbreak-nemotron\allowbreak-ultra\allowbreak-253b}
\cite{nvidia_nemotron_ultra}. These are  ``reasoning'' or ``thinking'' models that typically use chain of thought (CoT) and execute intermediary reasoning steps before outputting a response; they are designed for complex inference, coding, and problem-solving tasks. \texttt{o1}, \texttt{o3-mini}, \texttt{DeepSeek-R1}, and \texttt{Claude Opus 4} (with ``extended thinking'') make extensive use of explicit CoT reasoning \cite{openai_o1, openai_o3mini, deepseekai2025deepseekr1incentivizingreasoningcapability, anthropic_claude_opus4}.  \texttt{Gemini} is a sparse mixture-of-experts model that leverages a large context window for reasoning \cite{geminiteam2024gemini15unlockingmultimodal, geminiteam2025gemini25pushingfrontier}.   \texttt{Nemotron}, which is based on \texttt{llama-3 instruct} \cite{llama_team2024herd} and can be prompted to CoT, is optimized for hardware efficiency at inference \cite{nvidia_nemotron_ultra}. The decoding settings of the LLMs are given in the Supplemental Document.

The benchmarking results to follow use \texttt{Claude Opus 4} with the ``extended thinking'' option enabled and a maximum budget of 10,000 tokens per message. However, \texttt{Nemotron} was evaluated with ``detailed thinking'' disabled, because its internal reasoning traces were interleaved with the outputs in ways that were counterproductive (see Supplemental Document). 

As discussed in Section \ref{sec:config},  \texttt{o1} and \texttt{o3-mini} used the native Pydantic API call in OpenAI, but the structured outputs of the other models (at the entity extraction and component selection steps) were routed through \texttt{GPT-4o} for Pydantic validation.

\subsection{Evaluation criteria}
Each model performed the full PhIDO workflow for the 102 prompts, with 5 independent trials per prompt, resulting in a total of 510 trials per model. All trials were performed using the Interpreter agent in ``build-from-scratch'' mode without user intervention at any subsequent step. The trials were either completed successfully or terminated with the first fatal error. Each trial output was manually reviewed and assigned a single outcome code, corresponding to the earliest failure point or full-pipeline success. These outcome codes are listed in Table~\ref{tab:bench-criteria}. EE, CS, SG, PC, and L are failure modes, while S denotes success.

\begin{table}
\centering
\caption{Codes for classifying the outcomes in the PhIDO benchmark trials}
\label{tab:bench-criteria}
\begin{tabularx}{\linewidth}{clX}
\hline
\textbf{Code} ($r$) & \textbf{Stage} ($st$) & \textbf{Outcome Definition} \\ \hline
EE & Entity extraction & Interpreter: Failure to fully and correctly identify photonic components explicitly or implicitly described by the user query. Errors include omission of components, misinterpretation of descriptions, or generation of inaccurate initial schematics. This includes failures due to errors caused by GPT-4o-routed Pydantic validation. \\ 
CS & Component selection & Designer: Incorrect or incomplete mapping of identified components to devices in the PDK. \\ 
SG & Schematic generation & Designer: Invalid DOT schematic due to wrong connections, port mismatches, or structurally incorrect topology. \\ 
PC & Parameter configuration & Designer: User-specified parameters omitted, mis-assigned, or set to non-physical values in the netlist. \\ 
L & Layout & Layout: Invalid GDSII layout because of placement or routing. \\ 
S & Success & Entire workflow completes without error, yielding a valid schematic and GDSII layout. \\ \hline
\end{tabularx}
\end{table}

To quantify the performance, we define several metrics. First,  the \textit{absolute occurrence rate} for a given outcome code type \( r \) and model \( m \) is defined as
\begin{equation}
\text{Absolute Occurrence}_{r,m} = \frac{N_{r,m}}{N_{\text{total}}},
\end{equation}
where \( N_{r,m} \) is the number of trials for model \( m \) that resulted in the outcome code \( r \), and \( N_{\text{total}}\) is the total number of trials per model. This metric represents the proportion of total executions that terminated at a particular stage or succeeded the entire pipeline for each model. To evaluate stage-wise accuracy independently of upstream failures, we define the \textit{conditional occurrence rate} as the probability that an outcome \( r \), as listed in Table~\ref{tab:bench-criteria}, occurs at stage \( st \) given that the trial has reached that stage. This is expressed as  
\begin{equation}
\text{Conditional Occurrence}_{r,st,m} = \frac{N_{r\text{ at } st, m}}{ N_{\text{total}} - \sum_{i \in \text{prior}(st)} N_{r\text{ at } i, m}},
\end{equation}  
where \( N_{r\text{ at } st, m} \) is the number of times outcome \( r \) occurs at stage \( st \) for model \( m \), and \(\text{prior}(st)\) denotes all earlier stages in the pipeline. This formulation normalizes outcome frequency by the number of trials that reach a given stage. Because trials terminate after the first critical error, later stages are evaluated on smaller and potentially easier subsets of prompts, introducing attrition bias that may inflate conditional rates. To mitigate this, we benchmarked the layout stage independently using a fixed set of valid DOT schematics (see Supplemental Document); similar evaluations were not feasible for CS, SG, and PC due to their dependence on upstream prompt interpretations.

\section{ Results}
\subsection{Accuracy}

\begin{figure}
    \centering
    \includegraphics[width=1\linewidth]{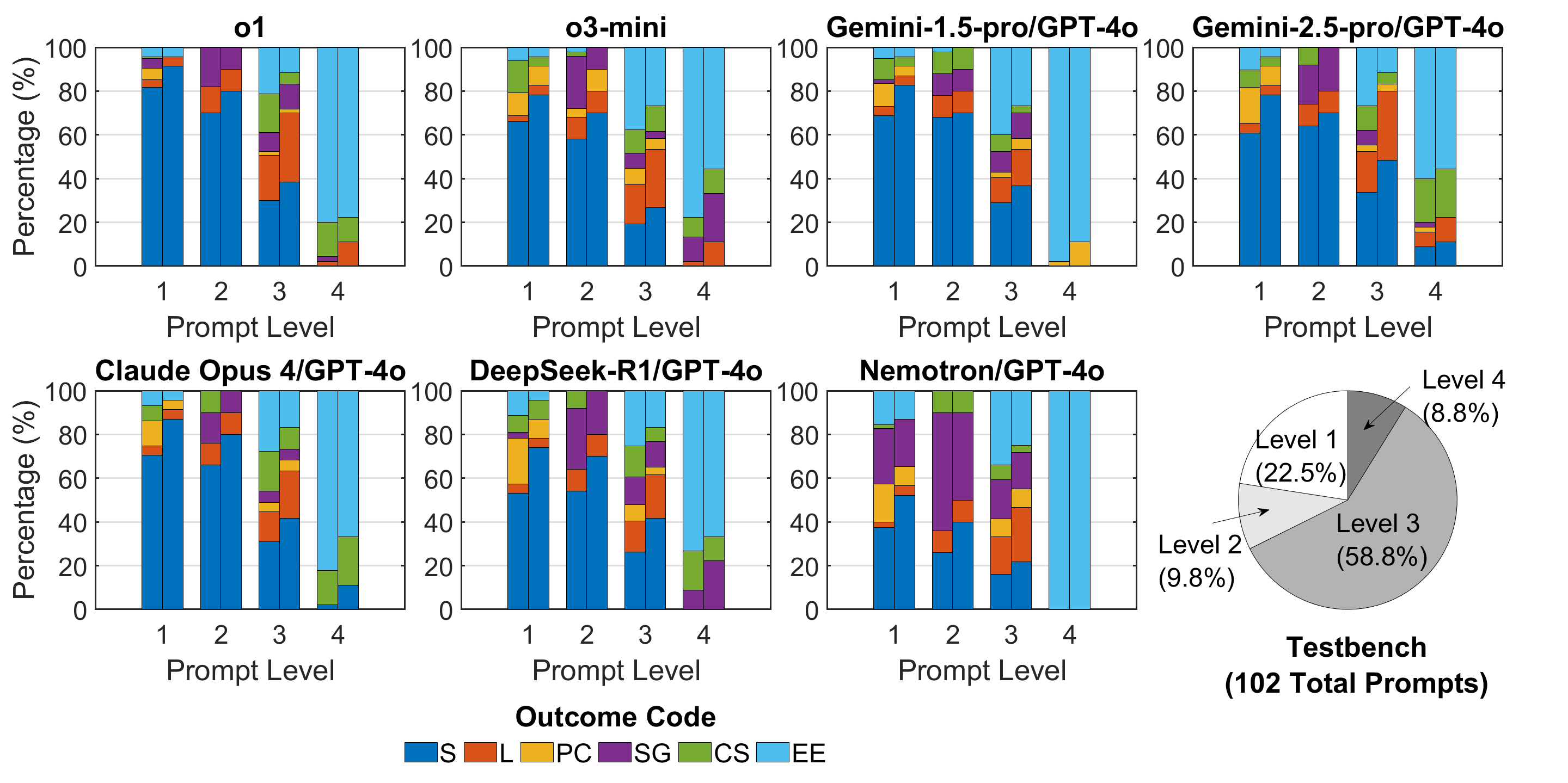}
    \caption{Absolute occurrence rate of each outcome category across 7 models, grouped by prompt complexity level. For each prompt level, the left bar represents pass@1 results and the right bar represents pass@5. Outcome codes are color-coded as shown in the legend below, with S (dark blue) indicating successful end-to-end execution. Non-OpenAI models were evaluated using structured outputs parsed through \texttt{GPT-4o} for schema validation via Pydantic. The pie chart summarizes the distribution of the 102 total prompts across the four complexity levels, with Level 3 comprising the majority of the testbench. The values are available in Table \ref{tab:outcome_pass1_pass5} of the Supplemental Document.}
    \label{fig:bench}
\end{figure}

\begin{figure}[h]
    \centering
    \includegraphics[width=1\linewidth]{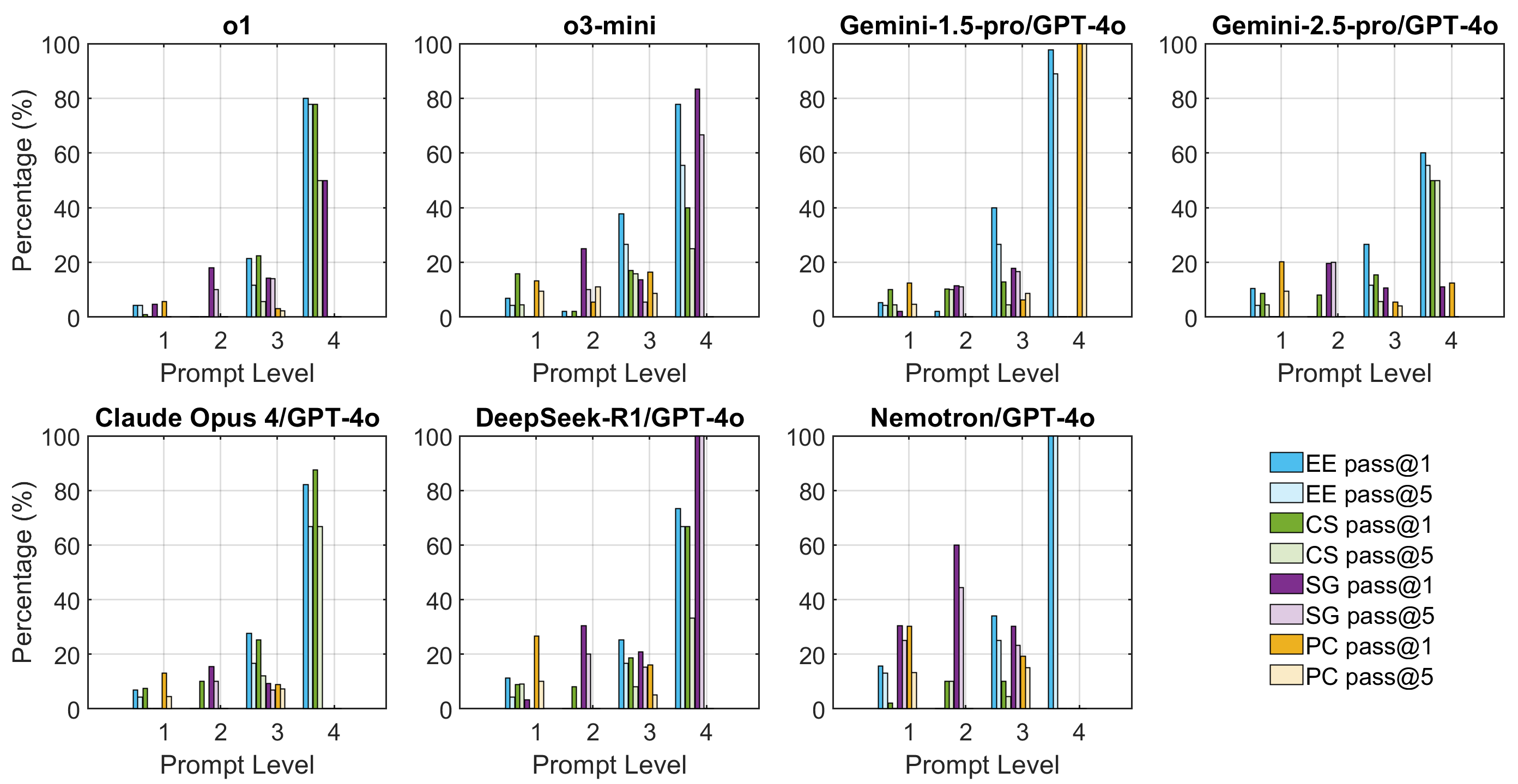}
    \caption{Conditional error rates for the EE, CS, SG, and PC stages across prompt levels for 7 language models. For non-OpenAI models, structured outputs were validated using \texttt{GPT-4o} with Pydantic. Due to upstream attrition, these rates may overestimate true performance and should be interpreted alongside Fig. \ref{fig:bias}. The values are available in Table \ref{tab:outcome_pass1_pass5_spreadsheet} of the Supplemental Document.}
    \label{fig:error}
\end{figure}

Figures~\ref{fig:bench} and~\ref{fig:error} summarize the absolute and conditional outcome statistics across all 7 models, reporting both pass@1 and pass@5 metrics~\cite{pass@metric}. The numerical values for these plots are available in the Supplemental Document. Although the LLMs were not tailored to photonic design, several models exhibited low stage-wise conditional error rates. These results should be interpreted in the context of attrition bias as discussed in Supplemental Document (Fig.~\ref{fig:bias}).

For Levels 1 to 3, which are low complexity, the top-performing models were \texttt{o1}, \texttt{Claude Opus 4}, and \texttt{Gemini-2.5-pro}, which achieved conditional success rates (i.e., (1 - Conditional Occurrence$_{r,st,m}$) for $r \ne S$) $>72\%$ at pass@1 and 80\% at pass@5 for all stages. Full end-to-end success rates were reduced due to the compounding difficulty of completing the entire workflow. The highest aggregate pass@1 success rates for Levels 1 to 3 were 47\% for \texttt{o1}, 45\% for \texttt{Claude Opus 4}, and 44\% for \texttt{Gemini-2.5-pro}. At pass@5, these increased to 58\% for \texttt{Gemini-2.5-pro}, 57\% for \texttt{Claude Opus 4}, and 56\% for \texttt{o1}. 

Levels 1 and 2 were handled well by \texttt{Gemini-1.5-pro}, \texttt{o1}, and \texttt{Claude Opus 4}, with pass@1 conditional success rates  $>93\%$ for EE, and $>82\%$ for the other stages. Most models maintained conditional success rates $> 70\%$, except \texttt{Nemotron}, which exhibited SG errors up to 60\% due to spurious components and missing connections in the schematic. PC errors at Level 1 were also high (20–30\%) for \texttt{DeepSeek-R1}, \texttt{Gemini-2.5-pro}, and \texttt{Nemotron}, caused by failures to extract parameters specified in the prompt such as MZI path length differences. An ablation study shows that enabling extended thinking in \texttt{Claude Opus 4} reduces SG and PC errors by over 10\% (see Supplemental Document).

By Level 3, all models showed a noticeable drop in performance, with end-to-end pass@5 success rates falling $<48\%$. The performance gap widened further at Level 4, where conditional error rates were $> 70\%$ for most models, and few trials advanced beyond EE. Level 4 includes structurally complex PIC topologies such as an $8\times 8$ Clements and Reck mesh and Spanke switches. Failures in CS at Level 4 were  common in large fanout circuits, such as a $1\times 16$ optical phased array containing more than 50 interconnected components, where models often selected  devices inconsistently (see Supplemental Document for an example). Among all LLMs, \texttt{Gemini-2.5-pro} was the most robust over all prompt complexities, maintaining conditional error rates $<60\%$ in all stages and successfully completing one Level 4 prompt four times. These results show much room exists for improvement in structural reasoning.

\subsection{Sources of errors}\label{sec:sourcesoferrors}

The EE stage was a dominant source of failure for all prompt complexities and models. As explained in the Methods section, the EE stage uses two types of LLMs: one that supports Pydantic schema validation and another that is schema-agnostic. EE errors fell into three main categories: (1) misinterpreting hierarchical structures; (2) incorrect enumeration, including missing or duplicated components; and (3) incorrect component connectivity. Figure \ref{fig:EE}(a) in the Supplemental Document compares the categorical breakdown of EE errors for two high-performing pipelines: one using \texttt{o1} end-to-end, and another combining \texttt{GPT-4o} with \texttt{Gemini-2.5-pro}. In the latter, 46\% of EE failures were traced to issues in the GPT-4o generated output, including incomplete component lists and hierarchy errors. Mixed model workflows can introduce fragility due to imperfect inter-model context transfer, where outputs from one LLM may not align with the assumptions or internal logic of the next, leading to  misinterpretations and logically incorrect circuits (see Supplemental Document).

Finally, we observed that some errors can be mitigated through sampling. Pass@5 consistently outperformed pass@1, highlighting the benefits of generating multiple completions. At Level 3, for example, \texttt{Gemini-2.5-pro} reduced its SG error rate by 11\% and EE error rate by 15\% under pass@5, while \texttt{o1} showed a 17\% reduction in CS errors. These improvements suggest that many failures are stochastic rather than systematic, often arising from output formatting inconsistencies or incomplete reasoning.

In addition to sampling, reducing the layout errors shown in Fig. \ref{fig:bench} would further improve end-to-end performance. To explore this, we conducted a preliminary investigation of improved routing and placement strategies, including component rotation. These modifications led to modest gains in schematic-to-layout success (see Supplemental Document). More advanced techniques, such as A* routing algorithms \cite{HartAstaralgo, arizonalidar}, may offer further improvements and remain a promising direction for future development.

\subsection{Runtime}

The runtime of end-to-end synthesis of designs using PhIDO is highly depending on the complexity level of the input prompt and the LLM used. Runtimes ranged from 40 seconds on a Level 1 complexity single component prompt using \texttt{gemini-1.5-pro} to  29 minutes on a Level 4 complexity prompt specifying a $1\times16$ optical phased array consisting of 63 components using \texttt{DeepSeek-R1}. However, the LLM used and average API provider latencies have a major impact on runtime. That same Level 4 complexity prompt was completed by \texttt{gemini-1.5-pro} in only 4 minutes and 35 seconds, though generated  layout was incorrect, unlike \texttt{DeepSeek-R1}. For the Level 1 prompt,  \texttt{DeepSeek-R1} generated the same correct  layout as \texttt{gemini-1.5-pro} but required more than five times the runtime (3 minutes and 33 seconds). Manual design by a human expert can range from minutes to several days, depending on circuit complexity, the designer’s expertise, and the level of rigor applied. Although a proof of concept, PhIDO illustrates AI-driven automation has the potential to significantly accelerate PIC design.

\subsection{Token usage}

\begin{sidewaystable*}[htbp]
\footnotesize
\centering
\caption{LLM Token Throughput \& Cost Comparison}
\label{tab:llm_token_cost}
\begin{tabular}{cc
    r@{\,}l  r@{\,}l  r@{\,}l  r@{\,}l  r@{\,}l  r@{\,}l  r@{\,}l}
\toprule
\textbf{Complexity} & \textbf{Metric} 
& \multicolumn{2}{c}{\texttt{o1} \footnotemark[1]}
& \multicolumn{2}{c}{\texttt{o3-mini} \footnotemark[1]}
& \multicolumn{2}{c}{\texttt{Gemini-1.5-pro}}
& \multicolumn{2}{c}{\texttt{Gemini-2.5-pro}}
& \multicolumn{2}{c}{\texttt{Claude Opus 4}}
& \multicolumn{2}{c}{\texttt{DeepSeek-R1}}
& \multicolumn{2}{c}{\texttt{Nemotron} \footnotemark[2]} \\
\textbf{Level} & & Mean & $\pm$ 95\%CI & Mean & $\pm$ 95\%CI & Mean & $\pm$ 95\%CI & Mean & $\pm$ 95\%CI & Mean & $\pm$ 95\%CI & Mean & $\pm$ 95\%CI & Mean & $\pm$ 95\%CI \\
\midrule

\multirow{4}{*}{1}
& Input Tokens     & 1537 & $\pm$ 31   & 1542 & $\pm$ 71   & 1632 & $\pm$ 57   & 1671 & $\pm$ 120  & 1913 & $\pm$ 26   & 1562 & $\pm$ 133  & 1688 & $\pm$ 126 \\
& Output Tokens    & 4706 & $\pm$ 619  & 3319 & $\pm$ 1226 & 305 & $\pm$ 50    & 345 & $\pm$ 116   & 1171 & $\pm$ 128   & 3208 & $\pm$ 825  & 372 & $\pm$ 78 \\
& Total Tokens     & 6243 & $\pm$ 633  & 4862 & $\pm$ 1286 & 1937 & $\pm$ 106  & 2015 & $\pm$ 235  & 3084 & $\pm$ 148  & 4770 & $\pm$ 880  & 2060 & $\pm$ 183 \\
& Total Cost (\$)  & 0.31 & $\pm$ 0.038 & 0.016 & $\pm$ 0.0055 & \bf 0.0036 & $\pm$ \bf 0.00032 & 0.0055 & $\pm$ 0.0013 & 0.12 & $\pm$ 0.010 & 0.0072 & $\pm$ 0.0018 & N/A\\
\midrule

\multirow{4}{*}{2}
& Input Tokens     & 1860 & $\pm$ 34   & 1811 & $\pm$ 22   & 1933 & $\pm$ 28   & 1987 & $\pm$ 86   & 2263 & $\pm$ 37 & 1830 & $\pm$ 14   & 1855 & $\pm$ 23 \\
& Output Tokens    & 7457 & $\pm$ 1082 & 4442 & $\pm$ 834  & 593 & $\pm$ 22    & 701 & $\pm$ 93    & 1840 & $\pm$ 191  & 6595 & $\pm$ 5296 & 568 & $\pm$ 20 \\
& Total Tokens     & 9317 & $\pm$ 1083 & 6253 & $\pm$ 840  & 2526 & $\pm$ 48   & 2688 & $\pm$ 173  & 4103 & $\pm$ 204 & 8425 & $\pm$ 5304 & 2423 & $\pm$ 32 \\
& Total Cost (\$)  & 0.48 & $\pm$ 0.065 & 0.022 & $\pm$ 0.0037 & \bf 0.0054 & $\pm$ \bf 0.00014 & 0.0095 & $\pm$ 0.001 & 0.17 & $\pm$ 0.015 & 0.015 & $\pm$ 0.012 & N/A \\
\midrule

\multirow{4}{*}{3}
& Input Tokens     & 2268 & $\pm$ 86   & 2255 & $\pm$ 49   & 2414 & $\pm$ 65   & 2503 & $\pm$ 118  & 2782 & $\pm$ 86  & 2275 & $\pm$ 42   & 2440 & $\pm$ 738 \\
& Output Tokens    & 9616 & $\pm$ 1621 & 6722 & $\pm$ 967  & 1059 & $\pm$ 64   & 1213 & $\pm$ 131  & 2891 & $\pm$ 412  & 6142 & $\pm$ 1383 & 1094 & $\pm$ 347 \\
& Total Tokens     & 11884 & $\pm$ 1666 & 8977 & $\pm$ 976  & 3472 & $\pm$ 127  & 3717 & $\pm$ 245  & 5673 & $\pm$ 470  & 8417 & $\pm$ 1383 & 3534 & $\pm$ 1084 \\
& Total Cost (\$)  & 0.61 & $\pm$ 0.099 & 0.032 & $\pm$ 0.0043 & \bf 0.0083 & $\pm$ \bf 0.0004 & 0.015 & $\pm$ 0.0015 & 0.26 & $\pm$ 0.032 & 0.014 & $\pm$ 0.003 & N/A \\
\midrule

\multirow{4}{*}{4}
& Input Tokens     & 4932 & $\pm$ 994  & 4160 & $\pm$ 791  & 5071 & $\pm$ 1507 & 4953 & $\pm$ 2067 & 6198 & $\pm$ 856 & 5078 & $\pm$ 360  & 6650 & $\pm$ 4425 \\
& Output Tokens    & 19450 & $\pm$ 7227 & 11677 & $\pm$ 4338 & 2915 & $\pm$ 641  & 3042 & $\pm$ 1169 & 8257 & $\pm$ 2754 & 20920 & $\pm$ 3011 & 4003 & $\pm$ 2778 \\
& Total Tokens     & 24382 & $\pm$ 8145 & 15837 & $\pm$ 5058 & 7986 & $\pm$ 2143 & 7995 & $\pm$ 3209 & 14454 & $\pm$ 3479 & 25998 & $\pm$ 2904 & 10653 & $\pm$ 7191 \\
& Total Cost (\$)  & 1.2 & $\pm$ 0.45 & 0.056 & $\pm$ 0.02 & \bf 0.021 & $\pm$ \bf 0.0051 & 0.037 & $\pm$ 0.014 & 0.71 & $\pm$ 0.22 & 0.047 & $\pm$ 0.0066 & N/A \\
\bottomrule
\end{tabular}
\footnotetext[1]{The reasoning token throughput of OpenAI reasoning models were not measured.}
\footnotetext[2]{Nemotron was accessed via NVidia NIM which did not have direct per-token pricing.}
\end{sidewaystable*}

Token analysis was conducted on a subset of the testbench comprising 20 prompts, selected evenly from the four complexity levels, with five prompts representing each level. 
Prompts exhibiting universally poor performance during performance benchmarking were excluded. Each selected prompt was executed for as many trials as needed to achieve five complete runs per LLM. Trials yielding invalid outputs causing unhandled exceptions were omitted from analysis. For each complexity level, the mean token throughput and corresponding standard deviations were computed. Standard deviations were derived by averaging the variances calculated individually across the five prompts within each complexity level. Subsequently, a 95\% CI was established using the t-distribution method with four degrees of freedom. The prompt specifying a 254 $\mu$m length straight waveguide proved particularly challenging for \texttt{Nemotron}, which was unable to complete any trial for this prompt due to incorrectly specifying the component as "\texttt{waveguide}" rather than using the correct PDK component "\texttt{straight}." CIs and mean throughput calculations were adjusted to account for the incomplete trials.

Token throughput was tracked using API-level metadata when available, as was the case with Google's and Anthropic’s models. For OpenAI models, the \texttt{o200k\_base} tokenizer was used, which captured only input and output tokens; consequently, tokens used internally during reasoning (hidden tokens) were not included in this analysis. \texttt{DeepSeek-R1} and \texttt{Nemotron} tokenizers were employed respectively to measure their specific token throughput. 

Cost estimates were derived from API provider. \texttt{DeepSeek-R1} and \texttt{Nemotron} are open-source models available on multiple inference service providers. We ran \texttt{DeepSeek-R1} on DeepSeek's  endpoint and Nvidia NIM, and we accessed \texttt{Nemotron} at Nvidia NIM. The \texttt{DeepSeek-R1} costs are quoted based on the DeepSeek endpoint. \texttt{Nemotron} was not included in the cost comparison because it was hosted on Nvidia NIM, where costs are based on graphical processing unit (GPU) usage time rather than the number of tokens. The effect of caching on cost was not analyzed, primarily because most tested models either do not implement automatic caching (\texttt{Claude Opus 4}) or do not expose caching-related metadata through their APIs (\texttt{o1, o3-mini, DeepSeek-R1}).

Table \ref{tab:llm_token_cost} summarizes the averages and corresponding 95\% confidence intervals (CI) for input, output, and total token counts, as well as the associated costs in US dollars as of August 5, 2025, for each LLM across the 4 prompt complexity levels. The lowest total cost for each input complexity level is in bold. As expected, the total token throughput required to execute user prompts increased  with the circuit complexity due to lengthier textual specifications.

The input token counts were similar across different models at each complexity level. This is likely due to regularization introduced at various stages of the PhIDO workflow. However, we observed significant differences in output token lengths and variability between models.  \texttt{o1}, \texttt{o3-mini}, \texttt{DeepSeek-R1}, and \texttt{Claude Opus 4} used about 3 to 15 times more output tokens than \texttt{Gemini-1.5-pro}, \texttt{Gemini-2.5-pro}, and \texttt{Nemotron} due to the explicit CoT, which also led to greater variability in the number of output tokens. 

CIs for token throughput were wider at higher complexity levels due to increased design variability. Prompts specifying more complex circuits were not necessarily more textually specific, which could lead to greater ambiguity. Additionally, the assignment of port-to-port connections is more challenging in higher complexity circuits. If waveguide crossings were detected, the Designer agent applied a self-correction step to iterate the circuit schematics (see Section \ref{sec:Designer}). This redesign process can significantly increase the variability in output token counts, as evident in Level 4 prompts.

Regarding costs, \texttt{o1} incurred the highest costs, because of its lengthy outputs and higher per-token pricing. \texttt{Claude Opus 4} was the most expensive model per token and the second most expensive model by average per-prompt cost. \texttt{DeepSeek-R1}, despite a high number of output tokens, cost less than \texttt{o1} and \texttt{o3-mini} at all complexity levels owing to its lower per-token rate. The \texttt{Gemini} models had the lowest costs.

\section{Discussion and Conclusion}


We have introduced PhIDO, a  multi-agent framework for PIC design automation that uses LLMs to generate GDSII layouts from natural language prompts. The endpoint in this work is primarily structural validity, rather than true fabrication sign-off using a commercial PDK and DRC deck. Unlike rule- or optimization-based methods tailored to specific design problems, PhIDO targets general-purpose PIC synthesis. We benchmarked 7 LLMs using 102 diverse user prompts. Since there exists some conceptual overlap between the user prompts and the corpus, the results should not be taken to reflect reasoning about PICs in isolation but rather the potential for agentic orchestration with retrieval. In this context,  we found that the state-of-the-art LLMs can perform well for simple PICs consisting of a few components, with \texttt{Gemini-2.5-pro} striking a balance between costs and accuracy in this study. However, the results may shift depending on the system prompts, schema validation approaches, and compute budgets (i.e., the lengths of the input, output, and hidden tokens, runtime).   

An important direction for future work is to improve the ability to correctly interpret vague or not fully specified inputs. Since many PIC descriptions, especially the hierarchy, component relationships, or parametric constraints, are not textual but rather captured in schematics, figures, tables, and equations, development of multi-modal models capable of joint reasoning over textual, mathematical, and visual technical data could increase PhIDO's accuracy in interpreting a prompt. To better translate design requests that use implicit assumptions and informal descriptors (e.g., ``MZI mesh'' or ``waveguide cutback structure'') into concrete topologies and component parameters, future work can include more rule-based reasoning and intermediate prompt refinement, or incorporate more iterative exchanges with the human user to sharpen specifications. 

Other directions for development include incorporating simulators/optimizers and expanding the photonics-specific knowledge base. Agentic frameworks, like PhIDO, can  smooth the adoption of accelerated solvers, topology optimization, and new design or simulation methods \cite{zhou2025intelligentelectronicphotonicdesignautomation,molesky2018inverse, SuSPINS2020, MacLellan2024, oquendo2025acceleratingphotonicintegratedcircuit, gehring2023femwell, oskooi2010meep, minkov2024gpu} to reduce design time and explore non-intuitive designs. As part of this integration,  the Designer and Verification modules could be integrated with an automated performance pass that invokes circuit and/or physical device simulations coupled to optimizers, so that a design can be ensured to meet user or automatically defined figures of merit (e.g., insertion loss, extinction ratio, bandwidth) before being classified as successful. Expanding the knowledge base to include more background knowledge, PDKs, and data on materials, process variation and yield will improve PhIDO's ability to interpret user queries and generate yield-aware designs and layouts for applications beyond silicon photonics.

Finally, PhIDO currently addresses only the design phase of PIC development; closing the loop with fabrication and experimental testing is essential for validating agent‐generated designs and autonomous PIC workflow. Future work can integrate automated measurement capabilities, using robotic wafer handlers, physics‐aware metrology, and model‐driven characterization, to realize self‐driving photonics laboratories with rapid design–fabrication–test cycles. 

Beyond the technical extensions, PhIDO and agentic frameworks for photonics can make PICs  more accessible to students and practitioners.  Achieving this vision will require rigorous standardization of agent interfaces, prompt schemas, knowledge representations, and data exchange protocols to ensure reliable, reproducible interactions across tools. As AI reasoning capabilities improve and integration with data sources, tools, and other agents become more standardized \cite{modelcontextprotocol, hou2025modelcontextprotocolmcp}, we anticipate that agents like PhIDO will evolve toward greater autonomy, accuracy, and domain specialization, ultimately enabling automated PIC development.

\section*{Funding}
This work was supported by the Natural Sciences and Engineering Research Council of Canada (Discovery grant, RGPIN-2024-05029) and the Max-Planck-Gesellschaft.

\section*{Acknowledgments}

J.K.S.P., J.M., and A.S. are thankful for fruitful discussions on AI-assisted photonics design with Troy Tamas of GDSFactory. J.K.S.P. and R. I. thank  Prof. Zongfu Yu, Prof. Shanhui Fan, Dr. Momchil Minkov, and Dr. Qing Hu at Flexcompute and Dr. Zuyang Liu at the University of Toronto for helpful discussions on Tidy3D. A.S. and J.K.S.P. thank Jason Liu, Cheick Doumbia, Onur Akdeniz and the Engineering Science ECE major capstone team from 2024 for initial investigations of LLMs for PIC design. A.S. and J.K.S.P. are appreciative of helpful discussions with Luis Calderon, Zijian Zhang, Dr. Hasitha Jayatilleka, and Dr. Neel Choksi. The authors are grateful for the donations of computational resources from Google Cloud, Nvidia, Flexcompute, and Axiomatic\_AI.

\section*{Disclosures}

D.R.E. is a director of Axiomatic\_AI, a company developing verifiable AI systems for science and engineering. J.P. is a scientific advisor to Axiomatic\_AI, and she is currently employed by Lightmatter, a company developing optical interconnects based on photonic integrated circuits. J.M. and V.A. are developing GDSFactory+, an extension of GDSFactory that provides advanced capabilities, including AI, for chip development. The other authors declare no competing interests.

\section*{Data and Code Availability} 

The source code for this work is available at the following Github repository  \url{https://github.com/JPPhotonics/PhIDO-Release}. The data from the benchmarking are located at \url{https://borealisdata.ca/dataverse/phido}. We welcome the community to build on this work.

\section*{Supplemental Document}
Please see the supplemental document.

\section*{AI Usage Statement}

AI coding assistants were  used to generate parts of the PhIDO software stack and documentation, which the authors subsequently verified. We employed LLMs as drafting aids for the manuscript. All LLM-generated prose and code were critically reviewed, edited, and validated by the human authors, who take full responsibility for the final content.

\section*{Author contributions}

A.S. designed and built the datasets and benchmarks, tested implementation approaches, analyzed the results, and guided junior team members in the project. Y.F. conducted benchmarking and token-analysis studies, analyzed the results, modularized the usability of the software, and created the repository for public release. V.A. implemented the initial codebase and conceived the PIC DSL. R.I. investigated and benchmarked the automation of FDTD simulations, and helped with the PDK in the knowledge base. F.K., K.M., S.A. and R.I.A. developed and benchmarked the layout agent. J.M. provided assistance with the GDSFactory integration. D.R.E. and J.K.S.P. conceived the project and AI methods for engineering. J.K.S.P. conceived the multi-agent PhIDO architecture, guided the overall direction, interpreted the data, and supervised the project. J.K.S.P., A.S., and Y.F. wrote the manuscript with inputs from the co-authors. 

\putbib

\end{bibunit}

\clearpage

    
 \pagebreak

\begin{bibunit}

\title{AI Agents for Photonic Integrated Circuit Design Automation: Supplemental document}

\setcounter{equation}{0}
\setcounter{figure}{0}
\setcounter{table}{0}
\setcounter{page}{1}
\setcounter{section}{0}

\makeatletter
\renewcommand{\theequation}{S\arabic{equation}}
\renewcommand{\thefigure}{S\arabic{figure}}
\renewcommand{\thesection}{S\arabic{section}}
\renewcommand{\thetable}{S\arabic{table}}
\renewcommand{\citeform}[1]{S#1}

\makeatletter
\renewcommand{\@biblabel}[1]{[S#1]}
\makeatother

\section{Decoding Settings of the Benchmarked LLMs}

Table \ref{tab:decoding} shows the decoding settings of the LLMs used in PhIDO.

\begin{table}[h]
\centering
\caption{LLM decoding settings for PhIDO benchmark.} \label{tab:decoding}
\small
\setlength{\tabcolsep}{6pt}
\begin{tabular}{lcccc}
\toprule
\textbf{Model} & \textbf{Temperature} & \textbf{Top-p} & \textbf{Seed} & \textbf{Max output tokens} \\
\midrule
\texttt{GPT-4o}               & 0.1  & —   & —   & 16384   \\
\texttt{o1}                   & N/A  & N/A  & —  & 100000   \\
\texttt{o3-mini}              & N/A  & N/A  & —   & 100000    \\
\texttt{Nemotron Ultra 253B}  & 0.1  & — & —    & 4096    \\
\texttt{Gemini-1.5-pro}       & 0.1  & —   & —   & 8192    \\
\texttt{Gemini-2.5-pro}       & 0.1   & —    & —   & 65535   \\
\texttt{Claude Opus 4}        & N/A  & —    & N/A  & 32000  \\
\texttt{DeepSeek R1}          & 0.6  & —    & —  &  32768  \\
\bottomrule
\end{tabular}
\caption*{\emph{N/A} denotes parameters not user-configurable for that model. “—” indicates parameters not explicitly set in the code, with provider defaults assumed.}
\end{table}

\section{Absolute occurrence rate values}
Table \ref{tab:outcome_pass1_pass5} shows the values of the absolute occurrence rates plotted in Fig. \ref{fig:bench} in the main manuscript.

\begin{sidewaystable*}[htbp]
\centering
\caption{Pass@1 and Pass@5 outcome absolute occurrence rates by complexity level and model}\label{tab:outcome_pass1_pass5}
\setlength{\tabcolsep}{4pt}
\renewcommand{\arraystretch}{1.1}
\begin{tabular}{cc*{7}{cc}}
\toprule
\multirow{2}{*}{Complexity} & \multirow{2}{*}{Outcome} &
\multicolumn{2}{c}{\texttt{o1}} &
\multicolumn{2}{c}{\texttt{o3-mini}} &
\multicolumn{2}{c}{\texttt{Gemini-1.5-pro}} &
\multicolumn{2}{c}{\texttt{Gemini-2.5-pro}} &
\multicolumn{2}{c}{\texttt{Claude Opus 4}} &
\multicolumn{2}{c}{\texttt{DeepSeek R1}} &
\multicolumn{2}{c}{\texttt{Nemotron}} \\
\cmidrule(lr){3-4}\cmidrule(lr){5-6}\cmidrule(lr){7-8}\cmidrule(lr){9-10}\cmidrule(lr){11-12}\cmidrule(lr){13-14}\cmidrule(lr){15-16}
 &  & p@1 & p@5 & p@1 & p@5 & p@1 & p@5 & p@1 & p@5 & p@1 & p@5 & p@1 & p@5 & p@1 & p@5 \\
\midrule
\multirow{6}{*}{1}
 & EE & 4.3 & 4.3 & 6.1 & 4.3 & 5.2 & 4.3 & 10.4 & 4.3 & 7.0 & 4.3 & 11.3 & 4.3 & 15.7 & 13.0 \\
 & CS & 0.9 & 0.0 & 14.8 & 4.3 & 9.6 & 4.3 & 7.8 & 4.3 & 7.0 & 0.0 & 7.8 & 8.7 & 1.7 & 0.0 \\
 & SG & 4.3 & 0.0 & 0.0 & 0.0 & 1.7 & 0.0 & 0.0 & 0.0 & 0.0 & 0.0 & 2.6 & 0.0 & 25.2 & 21.7 \\
 & PC & 5.2 & 0.0 & 10.4 & 8.7 & 10.4 & 4.3 & 16.5 & 8.7 & 11.3 & 4.3 & 20.9 & 8.7 & 17.4 & 8.7 \\
 & L  & 3.5 & 4.3 & 2.6 & 4.3 & 4.3 & 4.3 & 4.3 & 4.3 & 4.3 & 4.3 & 4.3 & 4.3 & 2.6 & 4.3 \\
 & S  & 81.7 & 91.3 & 66.1 & 78.3 & 68.7 & 82.6 & 60.9 & 78.3 & 70.4 & 87.0 & 53.0 & 73.9 & 37.4 & 52.2 \\
\midrule
\multirow{6}{*}{2}
 & EE & 0.0 & 0.0 & 2.0 & 0.0 & 2.0 & 0.0 & 0.0 & 0.0 & 0.0 & 0.0 & 0.0 & 0.0 & 0.0 & 0.0 \\
 & CS & 0.0 & 0.0 & 2.0 & 0.0 & 10.0 & 10.0 & 8.0 & 0.0 & 10.0 & 0.0 & 8.0 & 0.0 & 10.0 & 10.0 \\
 & SG & 18.0 & 10.0 & 24.0 & 10.0 & 10.0 & 10.0 & 18.0 & 20.0 & 14.0 & 10.0 & 28.0 & 20.0 & 54.0 & 40.0 \\
 & PC & 0.0 & 0.0 & 4.0 & 10.0 & 0.0 & 0.0 & 0.0 & 0.0 & 0.0 & 0.0 & 0.0 & 0.0 & 0.0 & 0.0 \\
 & L  & 12.0 & 10.0 & 10.0 & 10.0 & 10.0 & 10.0 & 10.0 & 10.0 & 10.0 & 10.0 & 10.0 & 10.0 & 10.0 & 10.0 \\
 & S  & 70.0 & 80.0 & 58.0 & 70.0 & 68.0 & 70.0 & 64.0 & 70.0 & 66.0 & 80.0 & 54.0 & 70.0 & 26.0 & 40.0 \\
\midrule
\multirow{6}{*}{3}
 & EE & 21.3 & 11.7 & 37.7 & 26.7 & 40.0 & 26.7 & 26.7 & 11.7 & 27.7 & 16.7 & 25.3 & 16.7 & 34.0 & 25.0 \\
 & CS & 17.7 & 5.0 & 10.7 & 11.7 & 7.7 & 3.3 & 11.3 & 5.0 & 18.3 & 10.0 & 14.0 & 6.7 & 6.7 & 3.3 \\
 & SG & 8.7 & 11.7 & 7.0 & 3.3 & 9.3 & 11.7 & 6.7 & 0.0 & 5.0 & 5.0 & 12.7 & 11.7 & 18.0 & 16.7 \\
 & PC & 1.7 & 1.7 & 7.3 & 5.0 & 2.7 & 5.0 & 3.0 & 3.3 & 4.3 & 5.0 & 7.7 & 3.3 & 8.0 & 8.3 \\
 & L  & 20.7 & 31.7 & 18.0 & 26.7 & 11.3 & 16.7 & 18.7 & 31.7 & 13.7 & 21.7 & 14.0 & 20.0 & 17.3 & 25.0 \\
 & S  & 30.0 & 38.3 & 19.3 & 26.7 & 29.0 & 36.7 & 33.7 & 48.3 & 31.0 & 41.7 & 26.3 & 41.7 & 16.0 & 21.7 \\
\midrule
\multirow{6}{*}{4}
 & EE & 80.0 & 77.8 & 77.8 & 55.6 & 97.8 & 88.9 & 60.0 & 55.6 & 82.2 & 66.7 & 73.3 & 66.7 & 100.0 & 100.0 \\
 & CS & 15.6 & 11.1 & 8.9 & 11.1 & 0.0 & 0.0 & 20.0 & 22.2 & 15.6 & 22.2 & 17.8 & 11.1 & 0.0 & 0.0 \\
 & SG & 2.2 & 0.0 & 11.1 & 22.2 & 0.0 & 0.0 & 2.2 & 0.0 & 0.0 & 0.0 & 8.9 & 22.2 & 0.0 & 0.0 \\
 & PC & 0.0 & 0.0 & 0.0 & 0.0 & 2.2 & 11.1 & 2.2 & 0.0 & 0.0 & 0.0 & 0.0 & 0.0 & 0.0 & 0.0 \\
 & L  & 2.2 & 11.1 & 2.2 & 11.1 & 0.0 & 0.0 & 6.7 & 11.1 & 0.0 & 0.0 & 0.0 & 0.0 & 0.0 & 0.0 \\
 & S  & 0.0 & 0.0 & 0.0 & 0.0 & 0.0 & 0.0 & 8.9 & 11.1 & 2.2 & 11.1 & 0.0 & 0.0 & 0.0 & 0.0 \\
\bottomrule
\end{tabular}

\end{sidewaystable*}

\section{Conditional occurrence rate values}
Table \ref{tab:outcome_pass1_pass5_spreadsheet} shows the conditional occurrence rates plotted in Fig. \ref{fig:error} in the main manuscript.

\begin{sidewaystable*}[htbp]
\caption{Pass@1 and Pass@5 conditional occurrence rates by complexity level and model}
\label{tab:outcome_pass1_pass5_spreadsheet}
\centering
\setlength{\tabcolsep}{4pt}
\renewcommand{\arraystretch}{1.1}
\begin{tabular}{cc*{7}{cc}}
\toprule
\multirow{2}{*}{Complexity} & \multirow{2}{*}{Outcome} &
\multicolumn{2}{c}{\texttt{o1}} &
\multicolumn{2}{c}{\texttt{o3-mini}} &
\multicolumn{2}{c}{\texttt{Gemini-1.5-pro}} &
\multicolumn{2}{c}{\texttt{Gemini-2.5-pro}} &
\multicolumn{2}{c}{\texttt{Claude Opus 4}} &
\multicolumn{2}{c}{\texttt{DeepSeek R1}} &
\multicolumn{2}{c}{\texttt{Nemotron}} \\
\cmidrule(lr){3-4}\cmidrule(lr){5-6}\cmidrule(lr){7-8}\cmidrule(lr){9-10}\cmidrule(lr){11-12}\cmidrule(lr){13-14}\cmidrule(lr){15-16}
 &  & p@1 & p@5 & p@1 & p@5 & p@1 & p@5 & p@1 & p@5 & p@1 & p@5 & p@1 & p@5 & p@1 & p@5 \\
\midrule
\multirow{4}{*}{1}
 & EE & 4.3 & 4.3 & 6.1 & 4.3 & 5.2 & 4.3 & 10.4 & 4.3 & 7.0 & 4.3 & 11.3 & 4.3 & 15.7 & 13.0 \\
 & CS & 0.9 & 0.0 & 15.7 & 4.5 & 10.1 & 4.5 & 8.7 & 4.5 & 7.5 & 0.0 & 8.8 & 9.1 & 2.1 & 0.0 \\
 & SG & 4.6 & 0.0 & 0.0 & 0.0 & 2.0 & 0.0 & 0.0 & 0.0 & 0.0 & 0.0 & 3.2 & 0.0 & 30.5 & 25.0 \\
 & PC & 5.8 & 0.0 & 13.2 & 9.5 & 12.5 & 4.8 & 20.2 & 9.5 & 13.1 & 4.5 & 26.7 & 10.0 & 30.3 & 13.3 \\
\midrule
\multirow{4}{*}{2}
 & EE & 0.0 & 0.0 & 2.0 & 0.0 & 2.0 & 0.0 & 0.0 & 0.0 & 0.0 & 0.0 & 0.0 & 0.0 & 0.0 & 0.0 \\
 & CS & 0.0 & 0.0 & 2.0 & 0.0 & 10.2 & 10.0 & 8.0 & 0.0 & 10.0 & 0.0 & 8.0 & 0.0 & 10.0 & 10.0 \\
 & SG & 18.0 & 10.0 & 25.0 & 10.0 & 11.4 & 11.1 & 19.6 & 20.0 & 15.6 & 10.0 & 30.4 & 20.0 & 60.0 & 44.4 \\
 & PC & 0.0 & 0.0 & 5.6 & 11.1 & 0.0 & 0.0 & 0.0 & 0.0 & 0.0 & 0.0 & 0.0 & 0.0 & 0.0 & 0.0 \\
\midrule
\multirow{4}{*}{3}
 & EE & 21.3 & 11.7 & 37.7 & 26.7 & 40.0 & 26.7 & 26.7 & 11.7 & 27.7 & 16.7 & 25.3 & 16.7 & 34.0 & 25.0 \\
 & CS & 22.5 & 5.7 & 17.1 & 15.9 & 12.8 & 4.5 & 15.5 & 5.7 & 25.3 & 12.0 & 18.8 & 8.0 & 10.1 & 4.4 \\
 & SG & 14.2 & 14.0 & 13.5 & 5.4 & 17.8 & 16.7 & 10.8 & 0.0 & 9.3 & 6.8 & 20.9 & 15.2 & 30.3 & 23.3 \\
 & PC & 3.2 & 2.3 & 16.4 & 8.6 & 6.2 & 8.6 & 5.4 & 4.0 & 8.8 & 7.3 & 16.0 & 5.1 & 19.4 & 15.2 \\
\midrule
\multirow{4}{*}{4}
 & EE & 80.0 & 77.8 & 77.8 & 55.6 & 97.8 & 88.9 & 60.0 & 55.6 & 82.2 & 66.7 & 73.3 & 66.7 & 100.0 & 100.0 \\
 & CS & 77.8 & 50.0 & 40.0 & 25.0 & 0.0 & 0.0 & 50.0 & 50.0 & 87.5 & 66.7 & 66.7 & 33.3 & N/A & N/A \\
 & SG & 50.0 & 0.0 & 83.3 & 66.7 & 0.0 & 0.0 & 11.1 & 0.0 & 0.0 & 0.0 & 100.0 & 100.0 & N/A & N/A \\
 & PC & 0.0 & 0.0 & 0.0 & 0.0 & 100.0 & 100.0 & 12.5 & 0.0 & 0.0 & 0.0 & N/A & N/A & N/A & N/A \\
\bottomrule
\end{tabular}

\end{sidewaystable*}

\section{Attrition bias}

Figure S1 illustrates the impact of attrition bias on stage-wise evaluation. At Level 4, most models failed early at the EE stage, sharply reducing the number of trials reaching CS, SG and PC. For example, while \texttt{Gemini-1.5-pro} demonstrated 0\% conditional error rates at the CS and SG stages for Level 4, this was based on a single surviving trial, limiting the reliability of the results.

\begin{figure}[h]
    \centering
    \includegraphics[width=1\linewidth]{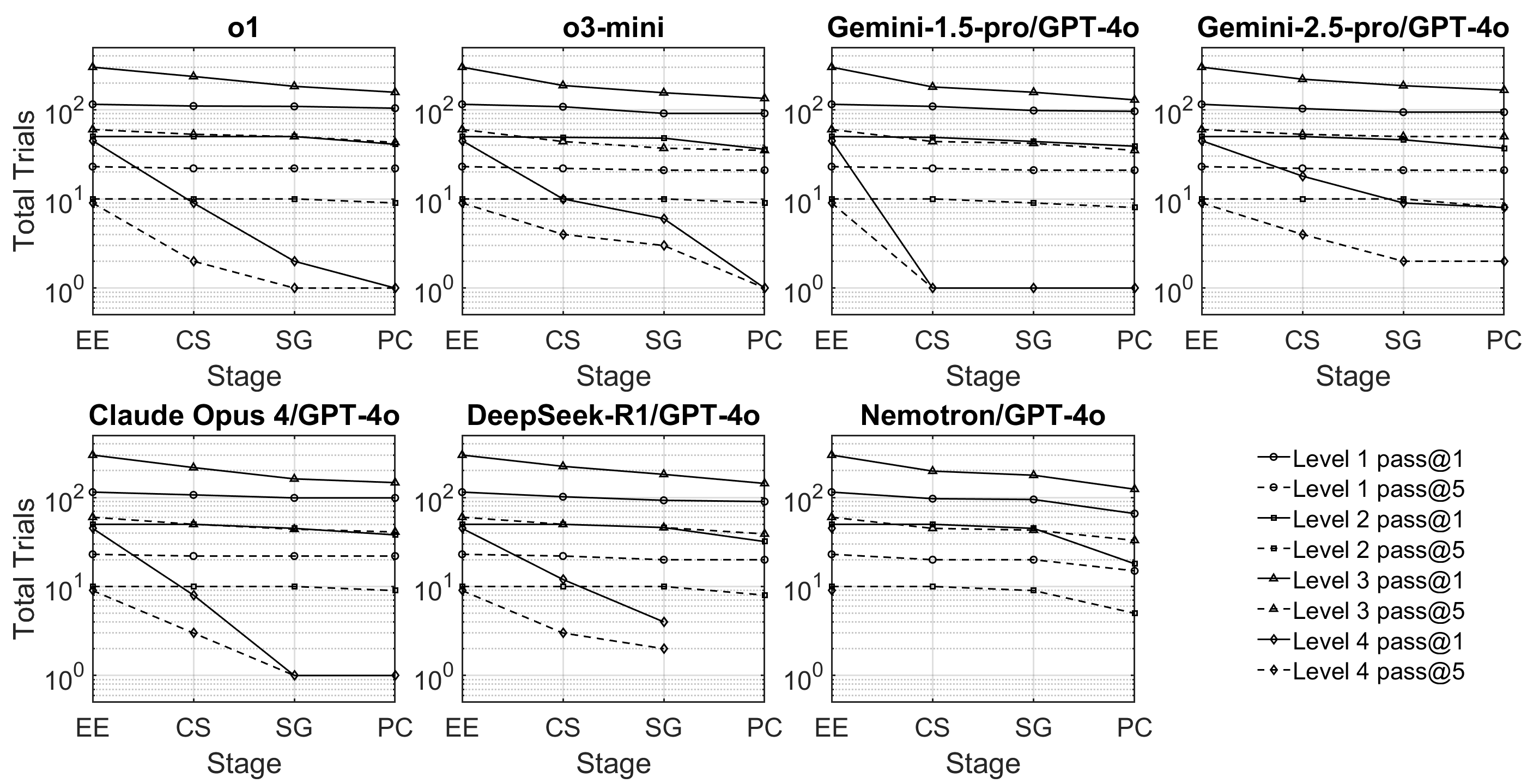}
    \caption{Attrition bias in the cross-model comparisons of stage-wise error rates.}
    \label{fig:bias}
\end{figure}

\section{Deep reasoning modes}

\texttt{Nemotron} and \texttt{Claude Opus 4} can be user-configured to run with or without deep reasoning (the option is called ``detailed thinking'' in \texttt{Nemotron} and ``extended thinking'' in \texttt{Claude Opus 4}.  Here, we compare the results with and without the deep reasoning feature.

\subsection{Nemotron}
The ``detailed thinking'' in \texttt{Nemotron} had to be disabled for PhIDO. During our preliminary tests, we found that \texttt{Nemotron} with ``detailed thinking'' enabled often appended interjections and inline comments with incorrect placements which would require manual preprocessing to clean, particularly in structured outputs such as circuit DSL YAMLs or DOT graphs constructors, which resulted in errors when fed into downstream PhIDO steps. 

For example, in prompt 1, ``Connect a 2x2 MZI with integrated thermal heaters to a directional coupler with a length of 125 um, dy of 100 um, and dx of 100 um'' Nemotron returned the following DOT graph output:

\begin{lstlisting}
graph graph_name_placeholder {
  rankdir=LR;
  node [shape=record];
  N1 [label="{{<o2> o2|<o1> o1} | N1: mzi_2x2_heater_tin_cband | {<o3> o3|<o4> o4}}"];
  N2 [label="{{<o2> o2|<o1> o1} | N2: _directional_coupler | {<o3> o3|<o4> o4}}"];
  N1:o3 -- N2:o1;
  N1:o4 -- N2:o2;
}
# Assuming N1's right-top port connects to N2's left-bottom port for minimal crossing
# Connecting N1's right-middle to N2's left-top, maintaining counter-clockwise order and minimizing crossings
\end{lstlisting}

In this example, the output containing inline comments located outside of the graphviz DOT graph constructor caused an unhandled exception when the output is automatically fed to the graphviz library. We observed that with ``detailed thinking'' off, \texttt{Nemotron} did not exhibit such unpredictable behavior and generally performed better in downstream PhIDO tasks. 

\subsection{Claude Opus 4}

\begin{figure}
    \centering
    \includegraphics[width=1\linewidth]{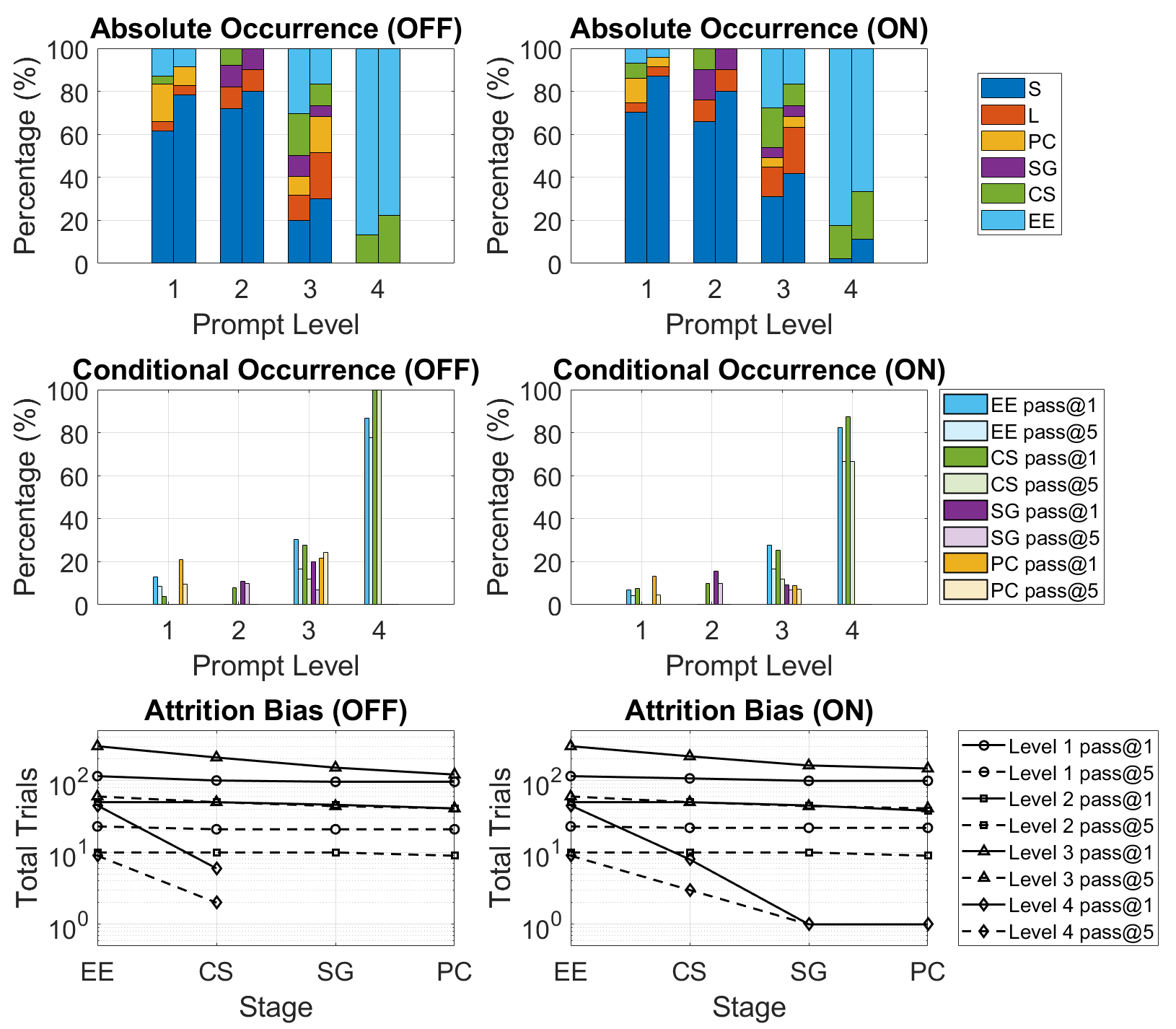}
    \caption{Ablation study of the deep reasoning feature in \texttt{Claude Opus 4}. Comparison of the absolute occurrence rate, conditional occurrence rate, and attrition bias with the ``extended thinking'' option off (left) or on (right) in \texttt{Claude Opus 4}.}
    \label{fig:ablationstudy}
\end{figure}

On the other hand, for \texttt{Claude Opus 4}, enabling  the ``extended thinking'' option significantly reduced conditional SG and PC error rates, as shown in Figure \ref{fig:ablationstudy}. At Level 1, PC error dropped from 21\% to 13\%. At Level 3, SG error fell from 20\% to 9\%, and PC error from 21\% to 9\%. These results suggest that deep reasoning improved the model's ability to parse numerical parameters and spatially reason over port connectivity when constructing schematics. 

\section{Example Stage-wise Errors}

\subsection{Entity Extraction}

In entity extraction, a common failure involved misinterpreting hierarchical descriptions. For example, flattening a Mach-Zehnder interferometer into a disjoint set of MMIs and phase shifters placed outside the MZI. This indicates the model’s difficulty in associating components with the larger structures they belong to. Additional errors included incorrect circuit connectivity and miscounting components (e.g., underestimating the number of input waveguides in a directional coupler). These findings suggest that while LLMs often correctly identify component types, they frequently struggle with descriptive modifiers and structural reasoning. Both are essential for generating correct circuit schematics.

\begin{figure}
    \centering
\includegraphics[width=1\linewidth]{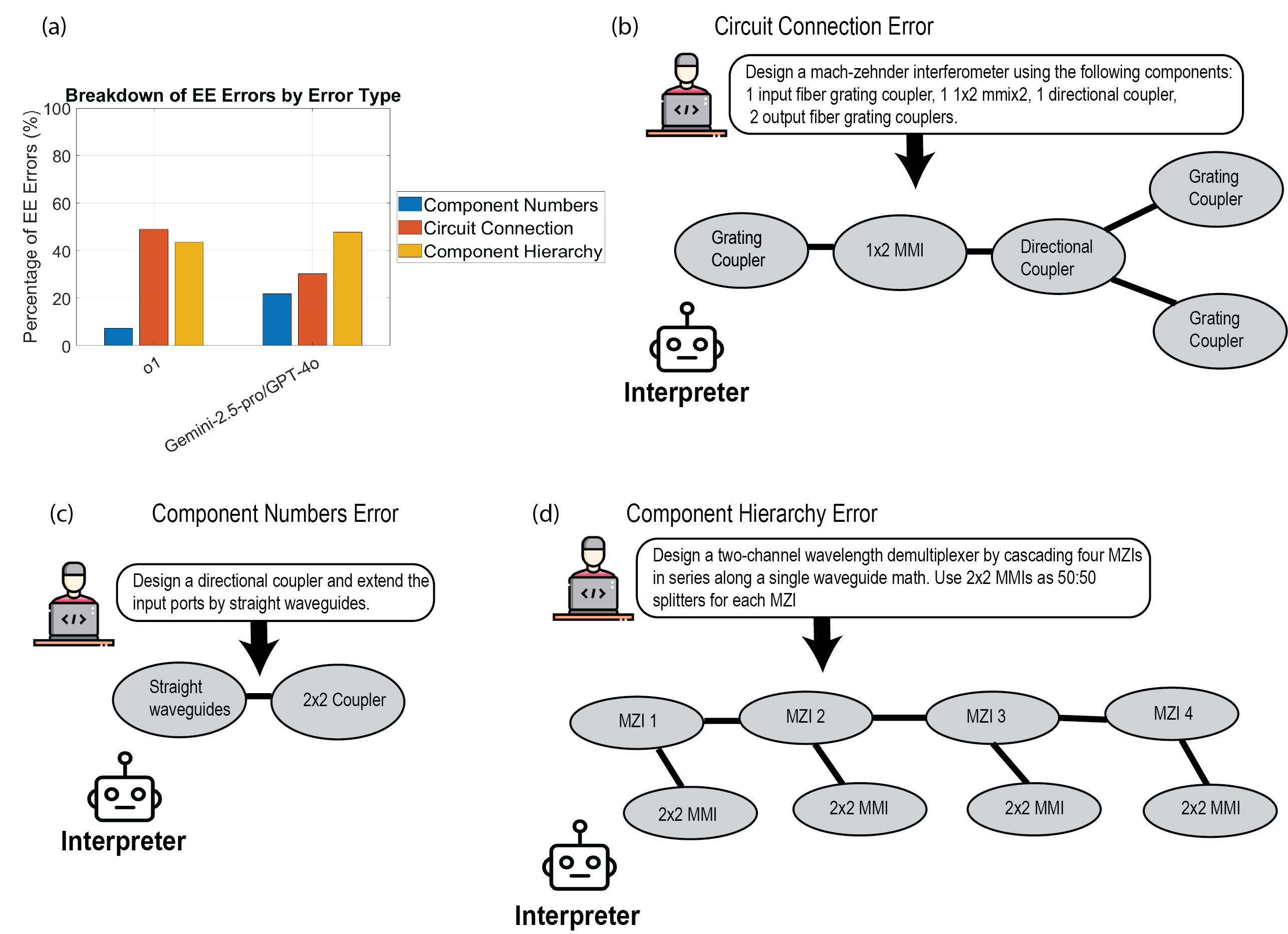}
    \caption{(a) Breakdown of entity extraction errors by error type. (b) Example of a circuit connection error. (c) An example of a component numbers error. (d) An example of a component hierarchy error. Here the descriptive modifiers of components are listed as separate components. }
    \label{fig:EE}
\end{figure} 

\subsection{Component Selection}

\begin{figure}
    \centering
    \includegraphics[width=1\linewidth]{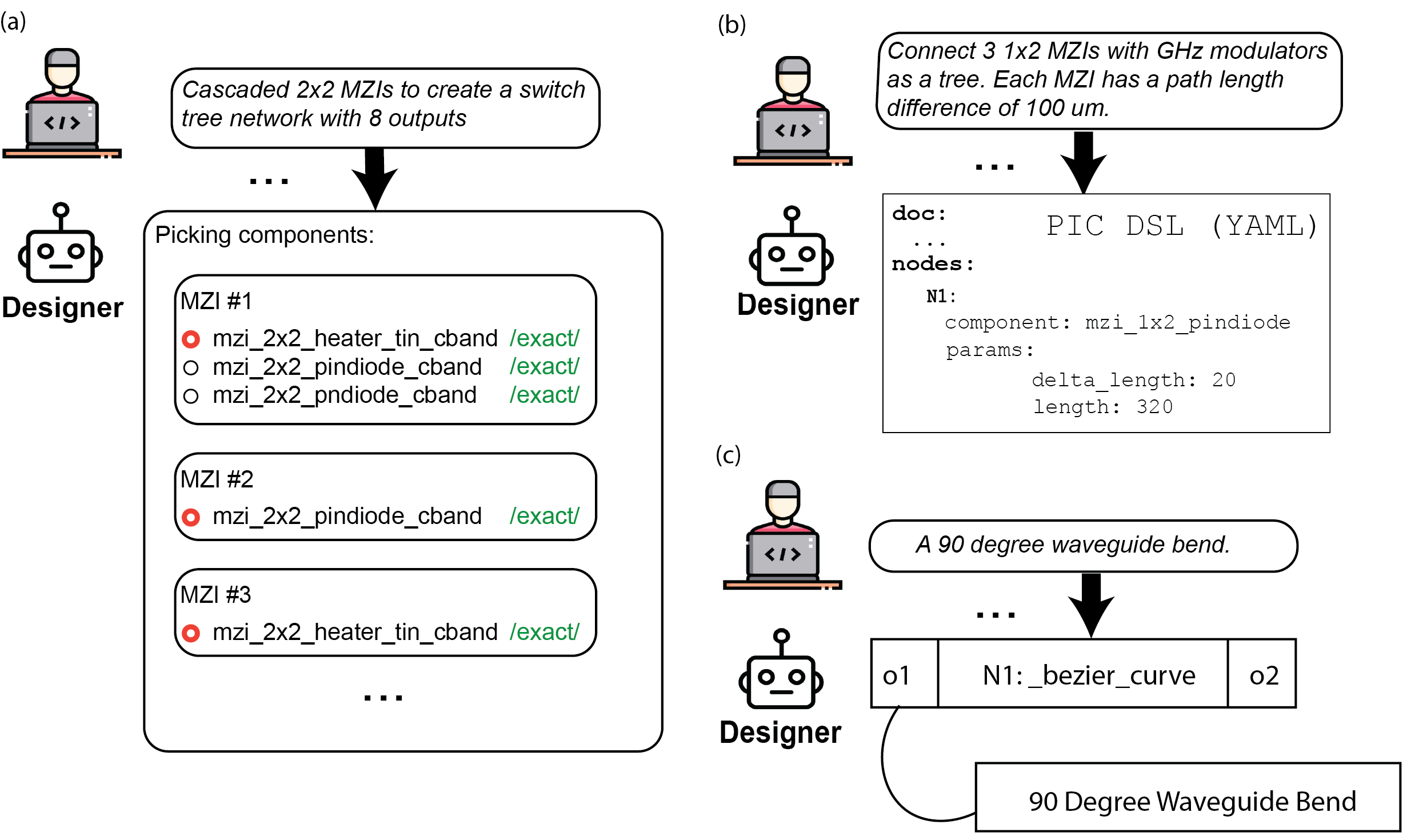}
    \caption{(a) Example of a component selection error. Although multiple $2\times2$ MZI variants exist in the PDK, the LLM occasionally retrieves only a partial or inconsistent set, resulting in devices with mismatched internal phase shifter types. (b) Example of a PCell configuration error in which the specified path length difference is not correctly parsed into the PIC DSL.(c) Example of a schematic generation error, where a component not present in the PDK is erroneously included in the schematic. }
    \label{fig:error_examples}
\end{figure}

A frequent cause of failure in component selection (CS) was the assignment of inconsistent devices to components that were intended to perform the same function, despite the clear requirement for uniform implementation (see Fig. \ref{fig:error_examples}). For example, the model might select a mix of modulator variants with different internal phase shifter types, such as thermo-optic and carrier-injection, when all instances should be identical. Parameter mismatches were also common, such as retrieving modulators with bandwidths that differed from user-specified values. While completion-based models allow partial control over output variability through temperature adjustments, reasoning agents have less exposure to such tuning because their behavior depends on internally managed multi-step planning and tool usage. Nevertheless, we observed these errors in GPT-4o, o1, and o3-mini, and they persisted in GPT-4o even when the temperature was set to 0.1.

\subsection{Schematic Generation}

An error frequently encountered in schematic generation is the addition of pseudo-components as nodes into the DOT graph schematic despite the PDK not containing such components. These errors arise from incorrect parsings of the PIC DSL by the LLM prior to DOT graph generation, as entities described in the DSL description or title in their respective fields are parsed as separate DOT graph nodes. For example, a DSL titled ``90 Degree Waveguide Bend'' containing a single \texttt{\_bezier\_curve} component is parsed into a DOT graph containing a bezier curve connected to a ``90 Degree Waveguide bend'' as a second component. We observed this error in most models, however the complexity levels at which this error in DSL parsing occurs varied according to model capability. Less performant models such as \texttt{Nemotron} with detailed thinking turned off can encounter this error even at level 1 complexity, whereas models such as \texttt{o1}, \texttt{Claude Opus 4}, or \texttt{gemini-2.5-pro} can encounter these errors at higher complexity level prompts. 

\subsection{PCell Configuration}

Specified component parameters in the user prompts may not be properly configured in the appropriate PCell field or at all (see Fig. \ref{fig:error_examples}). For example, a prompt specifying ``3 1x2 MZIs with GHz modulators as a tree. Each MZI has a path length difference of 100 $\mu$m'' may result in the generation of a DSL where selected MZI components do not have properly configured \texttt{params} and instead retain their default \texttt{delta\_length} value of 20 $\mu$m. PCell configuration issues are prominent with \texttt{DeepSeek-R1}, \texttt{Nemotron}, and \texttt{Gemini-2.5-pro}.

\section{Modifications to layout algorithm}

We explored improved routing and placement strategies to increase schematic-to-layout success. For test cases that could not be routed successfully using the original DOT-based placement and GDSfactory's river-router function, a brute-force rotation algorithm was applied, rotating each component in one of four orientations (north, east, south, and west) until a valid routing was found. The time complexity of this method is $4^{N}$, where $N$ is the number of components in the circuit, and was executed with a two minute runtime limit per circuit.

In PhIDO using GDSFactory v8.16, the baseline DOT-based routing successfully completed 74 out of 118 circuits, including 55 out of 72 smaller designs (\(\leq 5\) components) and 19 out of 46 larger designs (\(>5\) components). Adding the brute-force rotation step increased the number of successfully routed layouts to 90, with 67 successes for smaller designs and 23 for larger ones. Upgrading to GDSFactory v9.11.1  increased the number of successful routings from 94 without rotation to 109 with rotation.

Extending PhIDO’s capabilities to parse spatial layout constraints from natural language could further close the gap between schematic and physical design. For example, the prompt ``Design a single-bus ring resonator connected to two grating couplers for input/output coupling. The grating couplers should face away from each other and be vertically offset by 100 \textmu m'' encodes placement requirements. Parsing such constraints into the PIC DSL would enable more precise and controllable GDSII generation, ensuring that the final layout satisfies both functional and physical design intent.

\section{Mixed-model approach}
We also explored whether stage-wise ``best-of-breed'' approach could improve  performance by using a mixture of models where at each step of the workflow the LLM with the highest benchmarked conditional accuracy for that workflow step is used.
However, this strategy only yielded improvements at the entity extraction stage and did not improve the end-to-end success rate. For the stages subsequent to entity extraction, the performance either stagnated or degraded. Component selection performed worse than in any single-model workflow. This suggests that combining models in a stage-wise fashion negatively affects downstream performance.

The decline can be attributed to several compounding factors related to inter-model context transfer. First, different language models are sensitive to variations in prompt formatting and contextual framing. Outputs generated by one model may not be optimally structured for interpretation by the next, leading to subtle misalignments and degraded coherence. Second, each model encodes implicit assumptions and reasoning strategies that are not explicitly passed between stages. Switching models disrupts this continuity, forcing the downstream model to infer context without access to the upstream model’s internal logic. Finally, inconsistencies in knowledge representation, such as differences in parameter naming, component conventions, or domain-specific terminology, can cause semantic mismatches that may not trigger explicit errors but nonetheless compromise correctness. These effects are particularly pronounced in stages like component selection, where even minor contextual inconsistencies can lead to misinterpretation of component roles or invalid device retrieval. For example, DSL could not be successfully completed in several trials due to the generated YAML containing incorrect data fields because the component selection model tries unsuccessfully to match the entity extraction model's output with the YAML template. In summary, while model specialization appears promising, the lack of continuity in context and standardization in representation degrades the full-pipeline performance.

\end{bibunit}

\end{document}